\documentclass[aps,prd,twocolumn,showpacs,groupedaddress,nofootinbib,superscriptaddress]{
revtex4-1}
\usepackage{graphicx}
\usepackage{amssymb,amsmath,pifont}
\newcommand{\xmark}{\ding{55}}%

\begin{document}


\title{Interacting viscous dark fluids}




\author{Arturo Avelino}
\email[]{avelino@fisica.ugto.mx}
\affiliation{Departamento de F\'isica, DCI, Campus Le\'on,
Universidad de Guanajuato, \\
C\'odigo Postal 37150, Le\'on, Guanajuato, Mexico.}

\author{Yoelsy Leyva}
\email[]{yoelsy.leyva@fisica.ugto.mx}
\email[]{yoelsy.leyva@ucv.cl}
\affiliation{Departamento de F\'isica, DCI, Campus Le\'on,
Universidad de Guanajuato, \\
C\'odigo Postal 37150, Le\'on, Guanajuato, Mexico.}
\affiliation{Instituto de F\'isica, Pontificia Universidad
  Cat\'olica de Valpara\'iso\\
Casilla 4950, Valparaiso, Chile.}

\author{L. Arturo Ure\~na-L\'opez}
\email[]{lurena@fisica.ugto.mx}
\affiliation{Departamento de F\'isica, DCI, Campus Le\'on,
Universidad de Guanajuato, \\
C\'odigo Postal 37150, Le\'on, Guanajuato, Mexico.}

\date{\today}

\begin{abstract}
We revise the conditions for the physical viability of a cosmological
model in which dark matter has bulk viscosity and also interacts with
dark energy. We have also included radiation and baryonic matter
components; all matter components are represented by perfect fluids, except
the dark matter, that is treated as an imperfect fluid.
We impose upon the model the condition of a complete
cosmological dynamics that results in an either null or negative bulk
viscosity, but the latter also disagrees with the Local Second Law of
Thermodynamics. The model is also compared with cosmological
observations at different redshifts: type Ia supernova, the shift
parameter of CMB, the acoustic peak of BAO, and the Hubble parameter
$H(z)$. In general, observations consistently point out to a negative
value of the bulk viscous coefficient, and in overall the fitting
procedure shows no preference for the model over the standard
$\Lambda$CDM model.
\end{abstract}

\pacs{95.36.+x, 98.80.-k, 98.80.Es}
\maketitle


\section{Introduction}

Cosmological models with interacting dark components have gained
interest because it is expected that the most abundant components in the
present Universe, dark energy (DE) and dark matter (DM), interact with
each other, and some authors claim that some of these interaction
terms are promising mechanisms to solve the $\Lambda$CDM problems (see for
instance\cite{ChimentoJakubiPavon2000,*Copeland:2006wr,*Tsujikawa:2010sc,
Visco-Kremer2012}
and references therein).

On the other hand, it has been known since before the discovery of the present
accelerated expansion of the Universe that a bulk viscous fluid
may induce an accelerating cosmology 
\cite{HellerKlimek1975,*Barrow1986,*Visco-Padmanabhan1987,*Visco-Gron1990,
*Visco-Maartens1995,*Visco-Zimdahl1996}.
Hence, it has been proposed that the bulk viscous pressure can be one
of the possible mechanism to accelerate the Universe today (see
for instance
\cite{Cataldo:2005qh,Colistete:2007xi,*AvelinoUlises1P:2008,
AvelinoUlises2P:2010,
*Visco-RicaldiVeltenZimdahl2010,*Visco-AMontielNBreton2011}).
However, this idea still needs of some physically motivated model to
explain the origin of the bulk viscosity. In this sense some proposals
have been already put forward in\cite{Zimdahl:1999tn,*Mathews:2008hk}.

In the present work, we have the interest to explore and test an
interacting dark sector model which also takes into account a bulk
viscosity in the DM component. Our study is two fold: first, we
explore the general conditions for the model to have a complete
cosmological dynamics, and second, we use cosmological observations to
fit the free parameters of the model.

We have called {\it complete cosmological dynamics} to the fact that all
physically viable model must allow the existence of radiation and
matter domination eras at early enough times, so that the known
processes of the early Universe are not significantly changed with
respect to those of the standard Big Bang model. This seems to be an
usually overlooked condition in most studies of alternative
cosmological models, for which the primary concern is the present
accelerated expansion of the Universe, and then it is commonly thought
that a low-redshift analysis is quite enough for the task.

The full dynamics of the model is found through a dynamical system
analysis, a common tool in the analysis of cosmological
models\cite{Copeland:1997et,*Wainwright:1997,*Coley:1999uh,*UrenaLopez:2005zd,*Matos:2009hf,Caldera-Cabral2009,*Caldera-Cabral2010},
and then the DM-DE interaction term is chosen such as to allow the
writing of the equations of motion as an autonomous set of
differential equations. We are then able to write general conditions
for the existence of radiation and matter eras at early times that are
useful for a wide variety of interacting models.

The bulk viscous coefficient in our model is directly proportional to
the Hubble parameter, and we impose upon it a constraint that comes
from the Local Second Law of Thermodynamics (LSLT). In general, as it
also happens for our model, this latter condition selects only
positive definite values of the bulk viscous
coefficient\cite{Visco-IsraelStewart1979,*Visco-Israel1987,
*Visco-Maartens1996b,MisnerBook,*WeinbergBook}.

The model is also compared with different cosmological observations:
type Ia supernovae, the shift parameter of CMB, the acoustic peak of
BAO, and the Hubble parameter $H(z)$, in order to constraint its free
parameters. As we shall show, the fitted values acquire different
values depending on whether we use low-redshift or
intermediate-redshift observations. In a similar way as in the
condition for a complete cosmological dynamics, wrong conclusions may
be obtained if the analysis is only made with observations in the
lowest range of redshifts (late times).

The paper is organized as follows. In
Sec.~\ref{SectionDynamicalSystems} we present the full characteristics of
the model, the main equations of motion, and the dynamical system
analysis. The bulk viscosity of the model is represented by a single
free parameter, whereas the DM-DE interaction term is considered a
free function of the DM and DE density parameters, as long as the
dynamical system of equations remains autonomous. The cosmological
eras of the model are given in terms of the critical points of the
dynamical system, whose existence conditions depend upon the values of
the free parameters of the model. A detailed discussion about the
existence or not of appropriate cosmological eras is provided in terms
of the aforementioned constraint of a complete cosmological dynamics.

In Sec.~\ref{Case_F-equal-Alpha}, we focus our attention in a
particular form for the DM-DE interacting term that is directly
proportional to the DE energy density. Full details are given about
the existence and stability of the critical points, which are in turn
transformed into conditions upon the free parameters of the
model. Also, we show some particular examples of the dynamics of the
model for selected values of the free parameters.

We explain in Sec.~\ref{SectionCosmologicalObservations} the cosmological probes
that are used to constrain the model, and give separate examples of
the fitting procedure for different sub-cases of the model. For
completeness, we include here low and intermediate redshift
constraints, so that we can track the changes in the values of the
parameters for those cases. Finally, the main results are summarized
and discussed in Sec.~\ref{SectionConclusions}.

\section{Interacting bulk viscous dark fluids}
\label{SectionDynamicalSystems}
We study a cosmological model in a spatially flat
Friedmann-Robertson-Walker (FRW) metric, in which the matter
components are radiation, baryons, DM, and DE. Except to the DM, all
energy-matter components will be characterized by perfect fluids: radiation
and baryons are assumed to have the usual properties, whereas DM is treated
as an imperfect fluid having bulk viscosity, with a null hydrodynamical
pressure, and interacting with DE. This phenomenological model is a
natural extension of that proposed by Kremer and
Sobreiro\cite{Visco-Kremer2012}.

The Friedmann constraint and the conservation equations for the matter
fluids can be written as
\begin{subequations}
\label{eq:1} 
  \begin{eqnarray}
    H^2 &=& \frac{8 \pi G}{3}\left( \rho_{\rm r} + \rho_{\rm b} +
      \rho_{\rm dm} + \rho_{\rm de} \right) \,
    , \label{ConstrainFriedmann} \\
    \dot{\rho}_{\rm r} &=& - 4H\rho_{\rm r} \,
    , \label{ConsEqRadiation} \\
    \dot{\rho}_{\rm b} &=& - 3H\rho_{\rm b} \, , \label{ConsEqBaryon}
    \\
    \dot{\rho}_{\rm dm}  &=& - 3H \rho_{\rm dm} + Q -3H(-3H\zeta) \,
    , \label{EqConservationEffective4} \\
    \dot{\rho}_{\rm de} &=& - 3H\gamma_{\rm de}\rho_{\rm de} -Q
    \label{EqConservationEffective4a}
  \end{eqnarray}
\end{subequations}  
where $G$ is the Newton gravitational constant, $H$ the Hubble parameter,
$(\rho_{\rm r}, \rho_{\rm b}, \rho_{\rm dm}, \rho_{\rm de})$ are the energy
densities of the radiation, baryon, DM, and DE fluid components, respectively,
and $\gamma_{\rm de}$ is the barotropic index of the equation of state (EOS)
of DE, which is
defined from the relationship $p_{\rm de} = (\gamma_{\rm de} -1)
\rho_{\rm de}$, where  $p_{\rm de}$ is the pressure of DE.  The term
$-3H\zeta$ in Eq.~(\ref{EqConservationEffective4}) corresponds to the
bulk viscous pressure of the dark matter fluid, with $\zeta$ the bulk
viscous coefficient, whereas $Q$ is the DM-DE interaction term.

We consider the bulk viscous coefficient $\zeta$ as proportional to the total
matter density, $\rho_{\rm t} = \rho_{\rm r} + \rho_{\rm b} +
\rho_{\rm dm} + \rho_{\rm de}$, in the form
\begin{equation}
  \label{ViscosityDefinition}
  \zeta = \frac{\zeta_0}{\sqrt{24\pi G}} \rho^{1/2}_{\rm t} = \left(
\frac{1}{8 \pi G} \right) H \zeta_0\, ,
\end{equation}
where $\zeta_0$ is a dimensionless constant to be estimated from
the comparison with cosmological observations. From
Eq.~(\ref{ConstrainFriedmann}), we can see that this parametrization
corresponds to a bulk viscosity proportional to the expansion rate of
the Universe, i.e., to the Hubble parameter. Finally, the Raychadury
equation of the model is
\begin{equation}
  \dot{H} = -4\pi G \left( \frac{4}{3} \rho_{\rm r} + \rho_{\rm b} +
\rho_{\rm dm} + \gamma_{\rm de} \rho_{\rm de} -3H\zeta \right) \, .
\label{eq:R}
\end{equation}

In our analysis, we will take into account an important restriction
over the bulk viscous coefficient that comes from the Local Second Law
of Thermodynamics (LSLT). The \emph{local} entropy production for a
fluid on a FRW space--time is expressed
as\cite{WeinbergBook,MisnerBook}
\begin{equation}
  \label{entropy_definition}
  T \, \nabla_{\nu} s^{\nu} = \zeta (\nabla_{\nu} u^{\nu})^2 = 9H^2
  \zeta \, ,
\end{equation}
where $T$ is the temperature of the fluid, and $\nabla_{\nu} s^{\nu}$
is the rate of entropy production in a unit volume. Then, the second
law of the thermodynamics can be stated as $T \nabla_{\nu} s^{\nu}
\geq 0$; since the Hubble parameter $H$ is positive for an expanding
Universe, Eq.~(\ref{entropy_definition}) implies that $\zeta \geq
0$. For the present model, this inequality in turn becomes (see
Eq.~(\ref{ViscosityDefinition}))
\begin{equation}\label{entropy_condition}
\zeta_0 \geq 0 \, .
\end{equation}

\subsection{The dynamical system
  perspective \label{sec:dynam-syst-persp}}

In order to study all possible cosmological scenarios of the model, we
proceed to a dynamical system analysis of Eqs.~(\ref{eq:1})
and~(\ref{eq:R}). Let us first define the set of dimensionless
variables:
\begin{subequations}
\label{var}
\begin{eqnarray}
  x &=& \frac{8\pi G}{3H^{2}} \rho_{\rm de} \, , \quad y = \frac{8\pi
    G}{3H^{2}} \rho_{\rm dm} \, , \\
  u &=& \frac{8\pi G}{3H^{2}} \rho_{\rm b} \, , \quad z = \frac{8\pi
    G}{3H^{3}} Q \, .
\end{eqnarray}
\end{subequations}
Then, the equations of motion can be written in the following,
equivalent, form:
\begin{subequations}
\label{ev}
\begin{eqnarray}
  \frac{dx}{dN} &=& - z + x ( 4 - u - y - 3 \gamma_{\rm de} - 3 \zeta_0 )
  - x^2 ( 4 - 3\gamma_{\rm de} ) \, , \label{evx} \\
  \frac{dy}{dN} &=& y ( 1 - u - y - x ( 4 - 3\gamma_{\rm de} ) - 3\zeta_0
  ) + z + 3 \zeta_0 \, , \label{evy} \\
  \frac{du}{dN} &=& u ( 1 - u - y - x (4-3 \gamma_{\rm de} ) - 3\zeta_0
  ) \, , \label{evu}
\end{eqnarray}
\end{subequations}
where the derivatives are with respect to the $e$-folding number $N
\equiv \ln a$. In term of the new variables, the Friedmann
constraint~(\ref{ConstrainFriedmann}) can be written as:
\begin{equation}  
  \Omega_{\rm r} = \frac{8\pi G}{3H^2} \rho_{\rm r} = 1-x-y-u \, , \label{FCC}
\end{equation}
and then we can choose $(x,y,u)$ as the only independent dynamical
variables. 

Taking into account that $0\leq\Omega_{\rm r}\leq 1$, and imposing the
conditions that both the DM and DE components are both positive
definite and bounded at all times, we can define the phase space of
Eqs.~(\ref{ev}) as:
\begin{eqnarray}
  \Psi &=& \{(x,y,u): 0 \leq 1-x-y-u \leq 1 ,  0\leq x \leq1 \, ,
  \nonumber \\
  && 0 \leq y \leq 1, 0\leq u <1\} \, . \label{eq:space}
\end{eqnarray}
Other cosmological parameters of interest are the total effective EOS, 
$w_{eff}$, and the deceleration parameter, $q = - (1+\dot{H}/H^2)$,
which can be written, respectively, as
\begin{subequations}
\label{eq:2}
\begin{eqnarray}
  w_{eff} &=& \frac{1}{3} ( 1-u-y- x( 4- 3\gamma_{\rm de}) - 3\zeta_0 )
  \, , \label{effecw} \\
  q &=& \frac{1}{2}\{ 2-u-y-x (4- 3\gamma_{\rm de}) - 3\zeta_0 \} \,
  . \label{decel}
\end{eqnarray}
\end{subequations}

In order to obtain an autonomous system of ordinary differential
equations from Eqs.~(\ref{ev}), we will focus our attention hereafter
only in general interaction functions of the form $Q = 3H
f(\rho_{dm},\rho_{de})$ that can lead to closed functions
$z=z(x,y)$. As we shall see in the next section, this election will
allow us to impose general conditions over the variable $z$ (and on the
$Q$-term as well) in order to achieve a well behaved dynamics
(see\cite{Caldera-Cabral2009} for a similar exercise). Some examples
of the interaction $Q$ that lead to the desired form of $z$ are listed
in Table~\ref{tab1}.

\begin{table}[tp!]
\caption[Qf]{Some proposed forms of $Q(\rho_{de},\rho_{dm})$ for
  which the dynamical system~(\ref{ev}) becomes an autonomous system
  of differential equations.}
\begin{tabular}{c c c c}
  \hline \hline
  Model & $Q(\rho_{de},\rho_{dm})$ & $z(x,y)$ & References \\
  \hline
  i & $3H(\alpha_1 \rho_{de} + \alpha_2 \rho_{dm})$ & $3(\alpha_1 x +
  \alpha_2 y)$ &
  \cite{Quartin2008,Caldera-Cabral2009,*Caldera-Cabral2010} \\
  ii & $3H \lambda \frac{\rho_{de} \rho_{dm}}{\rho_{de} + \rho_{dm}}$
  & $3 \lambda \frac{xy}{x+y}$ & \cite{Zimdahl2003} \\
  iii & $3H \lambda \rho_{dm}$ & $3\lambda y$ & \cite{Zimdahl2001,
    Guo2007}\\
  \hline \hline
\end{tabular}\label{tab1}
\end{table}

\subsection{General conditions for a complete cosmological
  dynamics}\label{wellbehavecondition}
If the system of equations~(\ref{ev}) is autonomous, one then expects
that important stages in the evolution of the model be represented
by critical points in phase space. We will work on this hypothesis
here to make a description of the existence, or not, of the different
domination eras that have to be present in any model of physical
interest.

We then demand that our model must follow a complete cosmological
dynamics, namely: it should start in a radiation dominated era (RDE),
later enter into a matter dominated era (MDE), and finally enter into
the present stage of accelerated expansion; every one of these
statements can be translated in definite mathematical equations, that
we are going to discuss in detail in the sections below.

Before that, we need to calculate the critical points $(x_*, y_*,
u_*)$ of the dynamical system~(\ref{ev}), which are to be found from
the conditions:
\begin{subequations}
  \label{evo}
  \begin{eqnarray}
    0 & = & - z_* + x_{*} (4 - u_* - y_* - 3 \gamma_{de} - 3\zeta_0 ) -
    x_*^2 ( 4 - 3\gamma_{de}) \, , \label{evxo} \\
    0 &=& y_* ( 1 - u_* - y_* - x_* ( 4 - 3\gamma_{de} ) -
    3\zeta_0 ) + z_* + 3 \zeta_0 \, , \label{evyo} \\
    0 &=& u_* ( 1 - u_* - y_* - x_* ( 4-3 \gamma_{de} ) - 3\zeta_0
    ) \, , \label{evuo}
  \end{eqnarray}
\end{subequations}
where $z_* \equiv z(x_*,y_*,u_*)$ is the interaction variable
evaluated at the critical points, see Eqs.~(\ref{var}).

  \subsubsection{Radiation domination}

Let us start with the conditions for a purely RDE. According to the
Friedmann constraint~(\ref{FCC}), a purely RDE with $\Omega_r =1$
corresponds to $(x_*,y_*,u_*)=(0,0,0)$, and then Eqs.~(\ref{evo})
further dictate that
\begin{equation}
  z_* = 0 \, , \quad z_* = - 3 \zeta_0 \, . \label{ayr}
\end{equation}
The first condition on the DM-DE interaction term holds for many of
the interacting functions $z=z(x,y)$ in the specialized literature,
like for those examples  listed in Table \ref{tab1}; but the second
condition strongly implies that it is not possible to reconcile a
purely RDE with a non-zero bulk viscosity, $\zeta_0 \neq 0$.

However, there are other less extreme possibilities for radiation
domination in which a bulk viscosity exists, as long as we allow the
coexistence of radiation and other matter components early in the
evolution of the Universe. 

As the bulk viscosity term only appears actively for the equation of
motion of DM, see Eqs.~(\ref{evy}) and~(\ref{evyo}), we see that the
early presence of DM could allow the existence of bulk viscosity in a
RDE. The critical point we are looking for is of the form
$(x_*,y_*,u_*)=(0,y_*,0)$, under the assumption $y_* \ll 1$, and then
we obtain the following conditions,
\begin{equation}
  \label{eq:rde}
  y_* = - 3 \zeta_0 \, , \quad z_* = 0 \, .
\end{equation}
Thus, a RDE is possible as long as the DM-DE interacting term is
null, and the bulk viscosity is negative, $\zeta_0 < 0$ (in order to
preserve the condition $y \geq 0$). However, this is at variance with
the condition from the LSLT in Eq.~(\ref{entropy_condition}).

The null condition for the interaction term can be obtained if $z$ is
a function with mixed $x-y$ terms like that of Model (ii) in
Table~\ref{tab1}, or with a dependence only on $x$, an instance of
which is Model (i) with $\alpha_2=0$.

Another possible critical point for a RDE would be
$(x_*,y_*,u_*)=(x_*,0,0)$, which by means of Eqs.~(\ref{evo}), leads
to the conditions
\begin{equation}
  \label{eq:rde3}
     x_* = \frac{-3 \zeta_0}{4 - 3 \gamma_{de}} \, , \quad z_* = - 3
     \zeta_0 \, .
\end{equation}
As the DE EOS satisfies $\gamma_{de} < 1$, then a RDE is achieved if
$x_* \ll 1$ and $\zeta_0 < 0$, but the latter condition is again at
variance with the LSLT in Eq.~(\ref{entropy_condition}).

  \subsubsection{Matter domination} 

The existence of a MDE requires a scaling relation between the
baryonic and CDM densities in the form
$(x_*,y_*,u_*)=(0,\beta,1-\beta)$, where
$\beta\;\in\;[0,1]$\footnote{Only the values of $\beta$ in the range
  $[0,1]$ belong to the phase space~(\ref{eq:space}), and therefore
  make physical sense.}, so that $y_* + u_* =1$, as dictated by the
Friedmann constraint~(\ref{FCC}). This time, Eqs.~(\ref{evo}) dictate
that
\begin{equation}
     0 = -z_* \, , \quad 0 = - 3(1-\beta) \zeta_0 \, , \label{aum}
\end{equation}
are the simultaneous independent conditions to fulfill a MDE. 

The first condition requires again the interaction term $z$ to be a function
with mixed $x-y$ terms like that of Model (ii) in Table~\ref{tab1}, or
with a dependence only on $x$, like Model (i) with
$\alpha_2=0$. For this latter case, and also Model (iii), a nonzero
value of $\alpha_2$ needs a baryon dominated critical point,
$(x_*,y_*,u_*)=(0,0,1)$, which we consider as non-realistic.  

The second condition allows two possibilities:
\begin{itemize}
\item $\zeta_0=0$. As in the condition for a successful RDE, the model
  needs a null bulk viscosity to reach a correct MDE.
  
\item $\beta=1$ ($\forall\;\zeta_0\;\in\;[0,\infty)$) represents a
  critical point of pure CDM domination, which is at variance with the
  well established fact that baryons have a non negligible contribution
  to the matter contents.
\end{itemize}

Another scenario to describe the MDE is an scaling relation
among baryonic matter, CDM and DE. This requirement implies a fine
tunning over the very small amount of DE allowed for this period,
without preventing or slowing structure formation. This translates
into $(x_*,y_*,u_*)=(1 - y_* - u_*, y_*, u_*)$, so that $x_* + y_* +
u_* =1$, as indicated by the Friedmann constraint~(\ref{FCC}). With
the above values, Eqs.~(\ref{evo}) lead to two independent
possibilities:
\begin{itemize}
\item $z_* =3 (1 - y_*)((1 - \gamma_{de}) y_* - \zeta_{0})$, and
  $u_*=0$. The null contribution of baryons, and the scaling relation
  between CDM and DE, suggest that it is impossible to recover a
  successful MDE, even though this critical point could correspond to
  a possible late time scenario.
\item $z_* = 3 ( 1- \gamma_{de} )( 1- y_* - u_*)$, and  $x_* =
  \zeta_{0}/(\gamma_{de} -1)$. We have either: $\zeta_0 > 0$ and
  $\gamma_{de} > 1$, which agrees with the LSLT in
  Eq.~(\ref{entropy_condition}), but corresponds to a non realistic DE
  EOS, $w_{de}>0$; or $\zeta_0 < 0$ and $\gamma_{de} < 1$, which
  violates Eq.~(\ref{entropy_condition}), but somehow allows a valid
  MDE if $x_* \ll 1$.
\end{itemize}

  \subsubsection{Accelerated expansion}

In order to describe the present stage of accelerated expansion,
and at the same time alleviate the coincidence problem, we need a
scaling regime between the DM and DE components. This requirement
leads to the critical point $(x_*,y_*,u_*)=(x_*,1-x_*,0)$, and then
Eqs.~(\ref{evo}) lead to the single condition:
\begin{equation}
  z_* = 3 x_* ( 1 - x_* -\gamma_{de} + x_* \gamma_{de} - \zeta_0) \,
  . \label{axx}
\end{equation}
This last equation can be solved once the interaction term is given
for a particular model, and we can foresee that there must be valid
solutions of it for any values, positive or negative, of the bulk
viscosity constant $\zeta_0$. Moreover, if we impose the condition for
strict DE domination, $x_* = 1$, then $z_* = -3 \zeta_0$; this can be
possible, for instance, for Model (i) in Table~\ref{tab1}.

\subsubsection{Final comments}
The requirement of a complete cosmological dynamics discussed above,
from the dynamical system point of view, rules out any model that
obeys the equations of motion~(\ref{eq:1}), because the presence of
the bulk viscosity blockades the existence of standard RDE and MDE, if
we are to believe in the LSLT as stated in
Eq.~(\ref{entropy_condition}). It must be noticed, though, that an
accelerated expansion of the Universe at low redshifts is indeed
compatible with bulk viscosity.

In Secs.~\ref{Case_F-equal-Alpha}
and~\ref{SectionCosmologicalObservations} below, we will perform a
full dynamical system analysis of the field equations for the
particular case $Q= 3H \alpha \rho_{de}$, and then we will show the
importance of taking into account the full evolution of the Universe
to constraint cosmological models.

\section{The case for $Q=3H\alpha\rho_{de}$}
\label{Case_F-equal-Alpha}
\begin{center}
\begin{table*}[t!]\caption[Qf]{Location, existence conditions
    according to the physical phase space (\ref{eq:space}), and
    stability of the critical points of the autonomous system
    (\ref{evx})-(\ref{evu}) under $Q=3H\alpha\rho_{de}$. The
    eigenvalues of the linear perturbation matrix associated to each
    of the following critical points are displayed in
    Table~\ref{tab3}.}
  \begin{tabular}{@{\extracolsep{\fill}}| l | c | c | c | c | c |}
    \hline\hline
    $P_i$ & $x$ & $y$ & $u$ & Existence & Stability \\
    \hline\hline
    \footnotesize $1a$ & $0$ & $-3\zeta_0$ & $0$ & $-\frac{1}{3} \leq
    \zeta_{0} \leq 0$ & Unstable if $\zeta_{0}>-\frac{1}{3}$,
    $\alpha<\frac{4}{3}-\gamma_{de}$ \\
    \footnotesize & & & & & Saddle if $\zeta_{0}>-\frac{1}{3}$,
    $\alpha > \frac{4}{3} - \gamma_{de}$ or\\
    \hline
    \footnotesize $1b$ & $x$ & $-x-3 \zeta_0$ & $0$ & $\gamma_{de}=1$,
    $\alpha=\frac{1}{3}$ and ($\zeta_{0}=0$, $x=0$ or & \textit{Removed
      from phase space}\\
    \footnotesize & & & & $-\frac{1}{3} \leq \zeta_{0}<0$, $0\leq
    x\leq-3\zeta_{0}$) & \textit{See discussion in
      Sec.~\ref{sec:crit-points-stab}}\\
    \hline
    \footnotesize $1c$ & $x$ & \scriptsize$x(-4+3\gamma_{de})
    -3\zeta_{0}$ & $0$ & $\alpha = \frac{4}{3}- \gamma_{de}$,
    together with & Saddle
    if $\zeta_{0} < -\frac{1}{3}$\\
    & & & & those in Table~\ref{tab4} below. & \\
    \hline
    \footnotesize $2a$ & $0$ & $1$ & $0$ & \textit{Always} & Unstable
    if $\zeta_{0}<-\frac{1}{3}$, $\alpha<1-\gamma_{de}-\zeta_{0}$\\
    \footnotesize & & & & & Stable if $\zeta_{0}>0$, $\alpha>1 -
    \gamma_{de} -\zeta_{0}$ \\
    \footnotesize & & & & & Saddle if $\zeta_{0}<-\frac{1}{3}$, $\alpha >1
    -\gamma_{de}-\zeta_{0}$ or\\		
    \footnotesize & & & & & $-\frac{1}{3} < \zeta_{0}<0$, $\alpha \neq
    1-\gamma_{de}-\zeta_{0}$ or\\
    \footnotesize & & & & & $\zeta_{0} \geq 0$, $\alpha <
    1-\gamma_{de} -\zeta_{0}$\\
    \hline
    \footnotesize $2b$&$0$   &   $y$   &  $1-y$  & $\zeta_0=0$, $0<y\leq
    1$ & Saddle if $\alpha<1-\gamma_{de}$\\
    \hline 
    \footnotesize $2c$& $\frac{\zeta_0}{\gamma_{de}-1}$   &   $y$   &
    $1-y-\frac{\zeta_0}{\gamma_{de}-1}$  & $\alpha=1-\gamma_{de}$ and &
    Saddle if $\zeta_{0}>0$ \\
    \footnotesize   &   &   &   &( $\zeta_0>0$, $0\leq y< 1$,
    $\gamma_{de}\geq 1-\frac{\zeta_0}{y-1}$ or,&  \\
    \footnotesize   &   &   &   &  $\zeta_0<0$, $0\leq y< 1$,
    $\gamma_{de}\leq 1-\frac{\zeta_0}{y-1}$ or &  \\
    \footnotesize   &   &   &   & $\zeta_{0}=0$, $0<y\leq 1$,
    $\gamma_{de}\neq1$ ) &  \\
    \hline 
    \footnotesize $2d$& $x$   &   $y$   &  $1-x-y$  &
    $\zeta_0=\alpha=0$, $\gamma_{de}=1$ and &\textit{Removed from phase
      space}\\
    \footnotesize  &   &    &    &( $y=1$, $x=0$ or $y=0$, $0<x\leq 1$
    or &\textsl{See discussion in Sec.~\ref{sec:crit-points-stab}} \\
    \footnotesize  &   &    &    & $0<y<1$, $0\leq x \leq 1-y$ )&\\
    \hline
    \footnotesize $3a$&$1-\frac{\alpha+\zeta_0}{1-\gamma_{de}}$   &
    $\frac{\alpha+\zeta_0}{1-\gamma_{de}}$   &  $0$  &
    $\gamma_{de}<1$, $-\zeta_0\leq \alpha\leq1-\gamma_{de}-\zeta_{0}$
    or & Unstable if $\zeta_{0}>-\frac{1}{3}$,
    $\alpha>\frac{4}{3}-\gamma_{de}$ or  \\
    \footnotesize  &   &    &    & $\gamma_{de}>1$,
    $1-\gamma_{de}-\zeta_0\leq \alpha\leq -\zeta_{0}$&
    $\zeta_{0}\leq-\frac{1}{3}$, $\alpha>1-\gamma_{de}-\zeta_{0}$ \\
    \footnotesize  &   &   &   &   & Stable if $\zeta_{0}> 0$,
    $\alpha<1-\gamma_{de}-\zeta_{0}$ or\\
    \footnotesize  &   &   &   &   & $\zeta_{0}\leq 0$,
    $\alpha<1-\gamma_{de}$ \\
    \footnotesize  &   &   &   &   & Saddle if $\zeta_{0}>0$,
    $1-\gamma_{de}<\alpha<\frac{4}{3}-\gamma_{de}$ or\\
    \footnotesize  &   &   &   &   & $\zeta_{0}>0$,
    $1-\gamma_{de}-\zeta_{0}<\alpha<1-\gamma_{de}$ or\\
    \footnotesize  &   &   &   &   & $-\frac{1}{3}<\zeta_{0}\leq 0$,
    $1-\gamma_{de}-\zeta_{0}<\alpha<\frac{4}{3}-\gamma_{de}$ or\\
    \footnotesize  &   &   &   &   & $-\frac{1}{3}<\zeta_{0}<0$,
    $1-\gamma_{de}<\alpha<1-\gamma_{de}-\zeta_{0}$ or\\
    \footnotesize  &   &   &   &   & $\zeta_{0}\leq-\frac{1}{3}$,
    $1-\gamma_{de}<\alpha<\frac{4}{3}-\gamma_{de}$ or\\
    \footnotesize  &   &   &   &   & $\zeta_{0}<-\frac{1}{3}$,
    $\frac{4}{3}-\gamma_{de}<\alpha<1-\gamma_{de}-\zeta_{0}$\\
    \hline
    \footnotesize $3b$ &$1-y$   &   $y$   &  $0$  & $\alpha=-\zeta_0$,
    $\gamma_{de}=1$ & \textit{Removed from phase space}\\
     & & & & & \textit{See discussion in
       Sec.~\ref{sec:crit-points-stab}} \\
\hline \hline
\end{tabular}\label{tab2}
\end{table*}
\end{center}

\begin{table*}[t!]\caption[Qf]{Existence conditions for the critical
    point $P_{1c}$ according to the physical phase space
    (\ref{eq:space}).}
  \begin{tabular}{@{\extracolsep{\fill}}| c | c |}
    \hline \hline
    $P_i$ & $Existence$ \\
    \hline\hline
    $1c$   &  ($\zeta_{0}<-\frac{1}{3}$ and (
    $\gamma_{de}<\frac{4}{3}-\zeta_{0}$,
    $\frac{1+3\zeta_{0}}{-3+3\gamma_{de}}\leq x\leq
    \frac{3\zeta_{0}}{-4+3\gamma_{de}}$  or
    $\gamma_{de}=\frac{4}{3}-\zeta_{0}$,
    $x=\frac{3\zeta_{0}}{-4+3\gamma_{de}}$ )) or\\
    & ($\zeta_{0}=-\frac{1}{3}$ and ($\gamma_{de}<1$, $0\leq
    x\leq\frac{1}{4-3\gamma_{de}}$ or $\gamma_{de}>1$, $x=0$)) or\\
    & ($-\frac{1}{3}<\zeta_{0}<0$ and ($0\leq x\leq 1$,
    $\gamma_{de}=\frac{4}{3}+\zeta_{0}$ or $0\leq
    x\leq\frac{3\zeta_{0}}{-4+3\gamma_{de}}$,
    $1<\gamma_{de}<\frac{4}{3}+\zeta_{0}$ or $0\leq
    x\leq\frac{3\zeta_{0}}{-4+3\gamma_{de}}$, $\gamma_{de}<1$ or \\
    &  $0\leq x \leq \frac{1+3\zeta_{0}}{-3+3\gamma_{de}}$,
    $\gamma_{de}>\frac{4}{3}+\zeta_{0}$)) or \\
    & ($\zeta_{0}=0$ and ($x=0$, $\gamma_{de}<1$ or $x=0$,
    $1<\gamma_{de}<\frac{4}{3}$ or $0\leq x\leq 1$,
    $\gamma_{de}=\frac{4}{3}$ or $0\leq x \leq
    \frac{1}{-3+3\gamma_{de}}$ )) or\\
    & ($\zeta_{0}>0$ and ($x=1$, $\gamma_{de}=\frac{4}{3}$ or
    $\frac{3\zeta_{0}}{-4+3\gamma_{de}}\leq x \leq
    \frac{1+3\zeta_{0}}{-3+3\gamma_{de}}$,
    $\gamma_{de}>\frac{4}{3}+\zeta_{0}$))\\
    \hline \hline
\end{tabular}\label{tab4}
\end{table*}

\begin{table*}[t!]\caption[Qf]{Eigenvalues and some basic physical
    parameters for the critical points listed in Table~\ref{tab2}, see
    also Eqs.~(\ref{var}) and~(\ref{eq:2}).}
  \begin{tabular}{@{\extracolsep{\fill}}| c | c | c | c | c | c | c |}
    \hline \hline
    $P_i$&$\lambda_1$ & $\lambda_2$& $\lambda_3$ &$w_{eff}$ &
    $\Omega_r$ & $q$\\
    \hline\hline
    $1a$   &   $1$   &  $4-3\gamma_{de}-3\alpha$  & $1+3\zeta_0$ &
    $\frac{1}{3}$ &    $1+3\zeta_0$  & $1$\\
    \hline
    $1b$   &   $1$   &  $0$  & $1+3\zeta_0$ &  $\frac{1}{3}$ &
    $1+3\zeta_0$  & $1$\\
    \hline
    $1c$   &   $1$   &  $0$  & $1+3\zeta_0$ &  $\frac{1}{3}$ &
    $1-3x(\gamma_{de}-1)+3\zeta_0$  & $1$\\
    \hline
    $2a$   &   $-1-3\zeta_0$   &  $-3\zeta_0$  &
    $-3(-1+\gamma_{de}+\alpha+\zeta_0)$&  $-\zeta_0$ &    $0$  &
    $\frac{1}{2}(1-3\zeta_0)$ \\
    \hline
    $2b$   &   $-1$   &  $0$  & $-3(-1+\gamma_{de}+\alpha)$&  $0$ &
    $0$  & $\frac{1}{2}$ \\
    \hline
    $2c$   &   $-1$   &  $0$  & $3\zeta_0$&  $0$ &    $0$  &
    $\frac{1}{2}$ \\
    \hline
    $2d$   &   $-1$   &  $0$  & $0$&  $0$ &    $0$  & $\frac{1}{2}$ \\
    \hline
    \;\;$3a$\;\;   &   $3(-1+\gamma_{de}+\alpha)$   &
    $-4+3\gamma_{de}+3\alpha$ & $3(-1+\gamma_{de}+\alpha+\zeta_0)$ &
    $-1+\gamma_{de}+\alpha$ &    $0$  &
    $\frac{1}{2}(-2+3\gamma_{de}+3\alpha)$\\
    \hline 
    $3b$   &   $0$   &  $-1-3\zeta_0$  & $-3\zeta_0$&  $-\zeta_0$ &
    $0$  & $\frac{1}{2}(1-3\zeta_0)$ \\
    \hline \hline
\end{tabular}\label{tab3}
\end{table*}


This model of interaction was studied by\cite{Visco-Kremer2012} in
the context of interacting DM-DE with the presence of bulk
viscosity. The model can be recovered from Model (i) in
Table~\ref{tab1} with $\alpha_1 = - \sqrt{3} \zeta_0$ and $\alpha_2=
0$. Nonetheless, our study below generalizes the model
in\cite{Visco-Kremer2012} by taking a general interaction constant
$\alpha$, and two new components in the cosmic inventory: radiation
and baryonic matter. We will comment on the model
of\cite{Visco-Kremer2012} at the end of this section.

The selected $Q$-term leads to the following dimensionless interaction
variable $z$:
\begin{equation}
  z=3\alpha x \, . \label{eq:so}
\end{equation}
The nine critical points of the autonomous system~(\ref{ev}),
together with the interaction term in Eq.~(\ref{eq:so}), are
summarized in Table~\ref{tab2}, whereas details about their stability
and relevance for cosmology are given in Table~\ref{tab3}.

\subsection{Critical points and stability \label{sec:crit-points-stab}}
\begin{figure}[ht!]
\includegraphics[width=8cm]{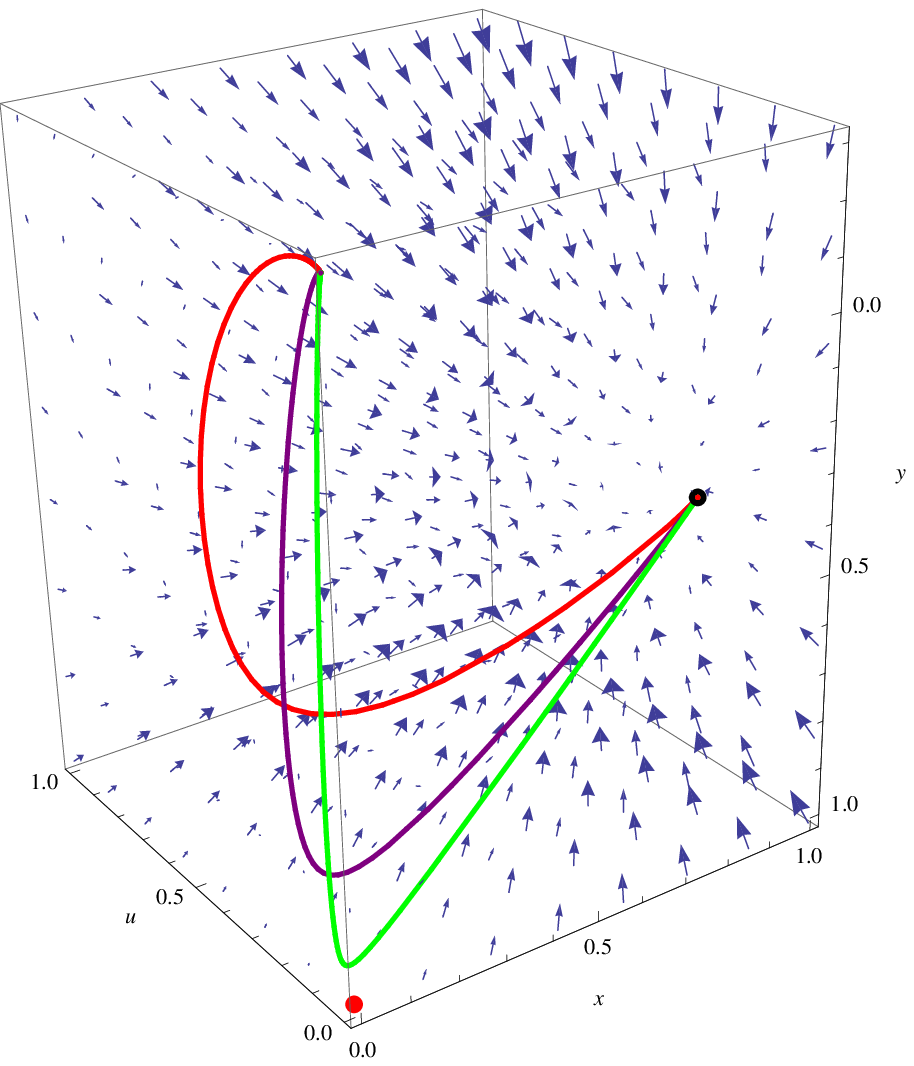}
\includegraphics[width=8cm]{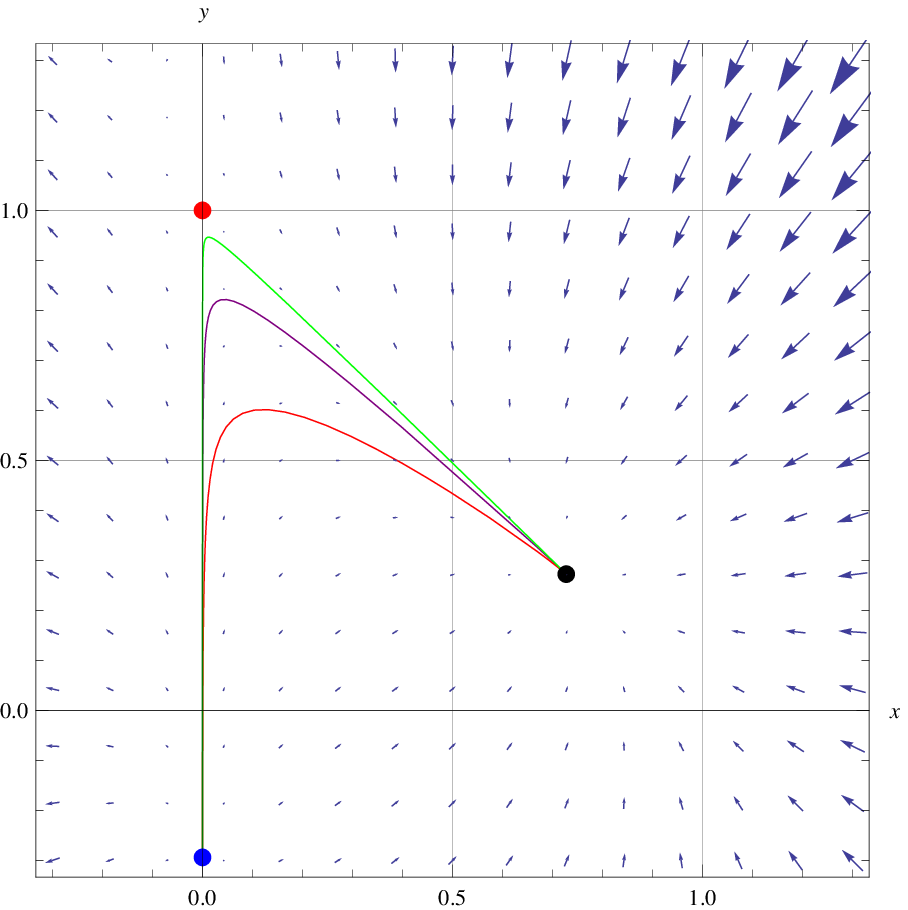}
\caption{\label{f1_2} Some orbits in the phase space for the choice
($\zeta_0$, $\gamma_{de}$, $\alpha$)=($0.098$, $0.2$, $0.12$). This
parameter election guarantee the saddle behavior of the pure DM
dominated solution $P_{2a}$ (red point) and the late time attractor
nature of $P_{3a}$, black point. Because of the nonzero value of
$\zeta_0$  the early time unstable solution corresponds to a
\textsl{wrong RDE} represented by blue point.} 
\end{figure}

The first point $P_{1a}$ corresponds to the co-existence of
radiation and DM, and exists if the bulk viscosity takes values in the
range $-\frac{1}{3} \leq \zeta_{0} \leq 0$. It also represents a
decelerating expansion solution with $q=1$ and $w_{eff}=1/3$. Critical
point $P_1$ exhibits two different stability behaviors
\begin{itemize}
\item Unstable if $\zeta_{0}>-\frac{1}{3}$ and $\alpha <
  \frac{4}{3}-\gamma_{de}$.
\item Saddle if $\zeta_{0}>-\frac{1}{3}$ and $\alpha > \frac{4}{3}-
  \gamma_{de}$.
\end{itemize}
In this point, the dimensionless energy parameter for radiation and DM
take the following values $\Omega_r=1+3\zeta_{0}$ and
$\Omega_{\rm dm} = -3\zeta_{0}$ respectively, as shown in
Table~\ref{tab3}. Therefore this point could represent a true RDE if
$\Omega_{\rm dm} \ll 1$, as long as $\zeta_{0}$ takes a negative value
very close to zero, or, in the most extreme case, if $\Omega_{\rm
  dm}=0$ then $\zeta_{0}=0$, meaning no bulk viscosity. In both cases,
the existence interval and the needed values for the bulk viscosity,
to archive a successful RDE, are outside the region of validity of the
LSLT ($\zeta_{0}>0$). \footnote{$\zeta_0>0$, as required by the LSLT,
  implies that for this critical point $y=\Omega_{\rm dm}<0$, and then
  we get a wrong RDE, see Fig.~\ref{f1_2}.}

The non hyperbolic critical point $P_{1b}$ exists if $\gamma_{de}=1$
and $\alpha=\frac{1}{3}$. The first one condition is at odds with our
expectation of a genuine DE fluid ($\gamma_{de} < 1$), and then we
will not take into account this critical point in our analysis.

$P_{1c}$ correspond to a decelerated solution ($q=1$) in which there
is radiation, DM, and DE. In effective terms, this point is able to
mimic the behavior of a radiation fluid ($w_{eff}=\frac{1}{3}$) but, a
truly RDE is only reached if $x\ll1$ and $-1\ll\zeta_{0}<0$, being the
latter condition in contradiction with the LSLT. If
$x=\frac{-3\zeta_{0}}{4-3\gamma_{de}}$, then this critical point
reproduces the analysis developed in the previous section, see
Eq.~(\ref{eq:rde3}). Despite its non-hyperbolic nature, $P_{1c}$
always has a saddle behavior if $\zeta_{0}<-\frac{1}{3}$, since it
possesses nonempty stable and unstable manifolds, see
Table~\ref{tab3}.

Critical point $P_{2a}$ represents a pure DM domination solution
($\Omega_{DM}=1$) and always exists, this fact is motivated by a null
contribution of baryonic matter. The stability of this fixed point is
the following:
\begin{itemize}
\item Saddle if $\zeta_{0}<-\frac{1}{3}$,
  $\alpha>1-\gamma_{de}-\zeta_{0}$ or \\ $-\frac{1}{3}<\zeta_{0}<0$,
  $\alpha\neq1-\gamma_{de}-\zeta_{0}$ or \\ $\zeta_{0}\geq 0$,
  $\alpha<1-\gamma_{de}-\zeta_{0}$
\item Stable if $\zeta_{0}>0$, $\alpha>1-\gamma_{de}-\zeta_{0}$.
\item Unstable if $\zeta_{0}<-\frac{1}{3}$,
  $\alpha<1-\gamma_{de}-\zeta_{0}$.
\end{itemize}

An interesting fact of $P_{2a}$ is the value of the effective EOS
parameter ($w_{eff}=-\zeta_{0}$): because of the non negative value of
the bulk viscosity constant required by the LSLT, $w_{eff}\leq 0$, which means that we
cannot recover a standard DM dominated picture, unless
$\zeta_0=0$. $P_{2a}$ is represented by a red point in
Fig.~\ref{f1_2}.

The non-hyperbolic fixed point $P_{2b}$ represents a scaling relation
between the baryonic and DM components. As we claimed before in
Sec.~\ref{SectionDynamicalSystems}, this critical point behaves as a
realistic MDE and exists only under a null bulk viscosity contribution
($\zeta_0=0$). If $\alpha<1-\gamma_{de}$ this critical point has a
saddle behavior.

$P_{2c}$ is a scaling solution between three components: baryons, DM,
and DE, and, unlike point $P_{2b}$, it exists for all values of
$\zeta_0$ (see Table~\ref{tab2} for the rest of existence
conditions). This critical point could represent a feasible MDE if
$x = \frac{\zeta_{0}}{\gamma_{de}-1} \ll 1$, and then  $0<\zeta_0\ll
1$. This implies a fine tunning over the bulk viscosity parameter due
to the negligible amount of DE that should exist during a MDE, which
would render it almost indistinguishable from $P_{2b}$ in the phase
space. This critical point exists given that:
\begin{itemize}
\item $\alpha=1-\gamma_{de}$, $\zeta_{0}>0$, $0\leq y<1$,
  $\gamma_{de}\geq 1-\frac{\zeta_{0}}{y-1}$. This region satisfies the
  LSLT (\ref{entropy_condition}), $\zeta_{de}>0$, but corresponds to a
  non truly DE component, $w_{de}>0$.
\item $\alpha=1-\gamma_{de}$, $\zeta_{0}<0$, $0\leq y<1$,
  $\gamma_{de}\leq 1-\frac{\zeta_{0}}{y-1}$. This region violates LSLT
  (\ref{entropy_condition}) but allow a valid MDE if the above
  condition, $0<\zeta_0\ll1$, is satisfied.
\item $\alpha=1-\gamma_{de}$, $0<y\leq 1$, $\gamma_{de}\neq 1$ and
  $\zeta_{0}=0$.
\end{itemize}
Despite of its non-hyperbolic nature, the critical point always
exhibits a saddle behavior if $\zeta_0 > 0$ since it has nonempty
stable and unstable manifolds.

Critical point $P_{2d}$ corresponds to a very particular selection of
the model parameters: $\alpha=\zeta_0=0$ and $\gamma_{de}=1$. These
values represent a model with a null interaction between DM and DE,
together with a null bulk viscosity contribution. The point $P_{2d}$
will not appear in the phase space as long as we take $\alpha \neq 0$
and $\zeta_0 \neq 0$.

Point $P_{3a}$ corresponds to a scaling solution between the DM and
DE components. From Table~\ref{tab2} we can note that this point
exists for any valid value of $\zeta_0$, and it represents an
accelerated solution if:
\begin{equation}
  \alpha < \frac{2}{3} - \gamma_{de} \, .
  \label{eq:acce}
\end{equation}
$P_{3a}$ exhibits an stable behavior if $\zeta_{0}> 0$,
$\alpha<1-\gamma_{de}-\zeta_{0}$, or $\zeta_{0}\leq 0$,
$\alpha<1-\gamma_{de}$. The first one condition is supported by the
LSLT, but the second is not. In the particular case $\alpha=-\zeta_{0}$
the strict DE domination is recovered ($\Omega_{de}=x=1$). The full
set of stability conditions for this critical point is shown in
Table~\ref{tab3}.

If $\gamma_{de}=1$, $\alpha=-\zeta_{0}$ and $\zeta_0> 0$, the critical
point $P_{3b}$ also appears in the phase space. However, the very
first condition is at variance with our expectations of a truly DE
fluid with $\gamma_{de} < 1$. Hence, this critical point will be
hereafter left out from our considerations.
\begin{figure}[ht!]
\includegraphics[width=8cm]{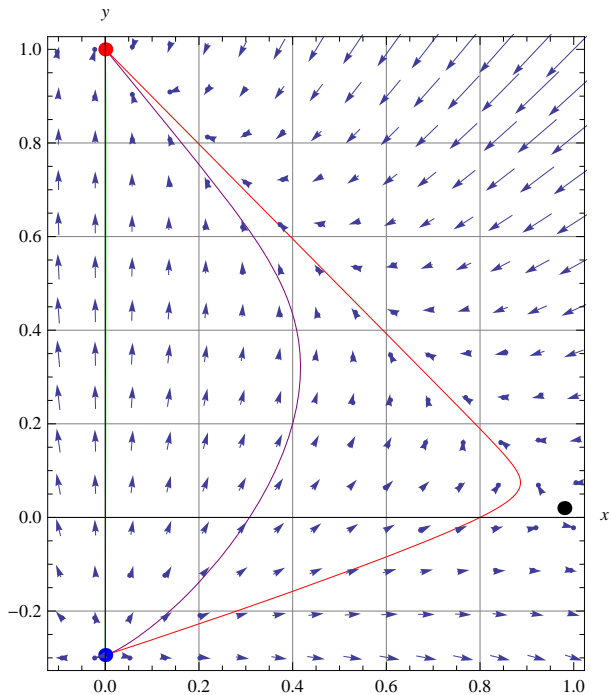}
\caption{\label{f_new} Some orbits in the phase space for the choice
($\zeta_0$, $\gamma_{de}$, $\alpha$)=($0.098$, $1.1$, $-0,1$). This
parameter election guarantee the existence of the saddle critical
point $P_{2c}$ and at the same time changes the dynamics of the phase
portrait: now the late time attractor is the DM dominated solution
$P_{2a}$ (red point) and the scaling solution between DM and DE,
$P_{3a}$ (black point), display a saddle type behavior. As
Table~\ref{tab2} shown, under this parameter choice $P_{3a}$ is
contained, as a particular case, in $P_{2c}$. Because of the nonzero
value of $\zeta_0$  the early time unstable solution corresponds to a
\textsl{wrong RDE} represented by the blue point.}
\end{figure}

\subsection{Cosmology evolution from critical
  points \label{sec:cosm-evol-from}}

According to our complete cosmological dynamics criterion, one of the
critical points of any physically viable model should correspond to a
RDE at early enough times, and this point should be an unstable point;
the unstable nature of this critical point guarantees that it can be
the source of any orbits in the phase space. The only two possible
candidates so far in our model are points $P_{1a}$ and $P_{1c}$. Both
cases require $-1\ll\zeta_{0}\leq0$ in order to be a true RDE
point, but such condition means a null contribution of bulk viscosity
($\zeta_{0}=0$), or else contradiction with the LSLT. Thus, we must
conclude that no critical point exists in the model that can represent
a RDE.

On the other hand, the evolution of the Universe requires the
existence of a long enough matter dominated epoch, in which DM and
baryons can be the dominant components. In our system, we need to look
carefully at critical points $P_2$ to search for an appropriate
candidate to be an unstable critical point dominated by the matter
components. 

In order to be in line with observations is better to avoid those
initial conditions that lead orbits to approach point $P_{2a}$, as it
does not permit the presence of baryons and its effective EOS is
negative, but it represents a point dominated solely by DM. 
Points  $P_{2b}$ and $P_{2d}$ must also be discarded, as
their existence always requires a null value of the viscosity
coefficient, and $P_{2d}$ even requires a null interaction between DM
and DE.

The only possibility seems to be point $P_{2c}$, as long as
observations could allow the presence of an early DE contribution to
the energy density of the Universe. In such a case, the value of the
viscosity coefficient $\zeta_0$ would have to be finely
adjusted. Unfortunately, as we showed in the previous discussion, this
critical point requires a non realistic DE component with EOS
$w_{de}>0$ ($\gamma_{de}>1$) in one case, and violation of the LSLT
through a negative value of the bulk viscosity ($\zeta_{0}<0$) in the
other.

Finally, we must get, as a possibility to alleviate the coincidence
problem of DE, a scaling solution with a nearly constant ratio
between the energy densities of DM and DE at late times, which should
in turn correspond to a stable critical point; the only one at hand  in
our system that could fulfill those expectations is $P_{3a}$. For the
allowed values of $\zeta_0$, this point represents a scaling solution
between the DE and DM components in the existence regions, and also
admits a pure DE domination solution if only $\alpha =-\zeta_{0}$
($\gamma \neq 1$). The required presence of the bulk viscosity limits
the possibility of choosing initial condition that lead orbits to
connect MDE to DM-DE scaling solution to the following possibilities:
\begin{itemize}
 \item Orbits that connect $P_{2a}$ with $P_{3a}$. Despite the stable
   and accelerated nature of the scaling solution $P_{3a}$, it is not
   possible to recover the RDE and MDE as previously discussed. In
   Fig.~\ref{f1_2} are shown some numerical integration of the
   autonomous system (\ref{evx}-\ref{evu}), for the interaction
   function (\ref{eq:so}) with ($\zeta_0$, $\gamma_{de}$,
   $\alpha$)=($0.098$, $0.2$, $0.12$). The orbits reveal that the
   $P_{3a}$ solution is the future attractor whereas the
   \textsl{wrong} RDE ($P_{1}$) is the past attractor. 
 
 \item Orbits that connect $P_{2c}$ with $P_{3a}$. The existence
   conditions of both critical points (see Table \ref{tab2}) also
   implies that DM-DE scaling solution mimics the behavior of
   pressureless matter ($w_{eff}=0$). In the same region, this
   solution is not accelerated ($q=\frac{1}{2}$) being impossible to
   explain the late-time behavior of the Universe. This result rules
   out those initial conditions leading to orbits connecting both
   critical points. Fig.~\ref{f_new} shows some example orbits in the
   $x-y$ plane.
\end{itemize}

Unlike the above cases, the presence of non-null bulk viscosity
entails no problem for a successful late-time accelerated evolution of
the Universe but is impossible to recover a well behaved picture of
the whole history of the Universe without be at variance with the
LSLT. The simultaneous presence of interaction between DM and DE and
bulk viscosity results in a very restrictive condition for the
model. 


\section{Cosmological
  constraints}\label{SectionCosmologicalObservations}
We now proceed to constrain the values of $(\zeta_0, \gamma_{\rm de},
\alpha)$, compute their confidence intervals, and calculate their best
estimated values, as we compare with different cosmological
observations that measure the expansion history of the Universe. For
future reference, we write here an explicit expression for the
normalized Hubble parameter, which is an exact result for the
model~(\ref{eq:1}):
\begin{multline}
E^2(z)  = \Omega_{\rm r0} (1+z)^4 + \Omega_{\rm b0} (1+z)^3 + \\
+ \Omega_{\rm de0} (1+z)^{3 ( \gamma_{\rm de} + \alpha)} +
\hat{\Omega}_{\rm dm}(z) \, ,
\end{multline}
where $E(z) \equiv H(z)/H_0$, and the cumbersome formula for
$\hat{\Omega}_{\rm dm}(z)$ is given in
Eq.~(\ref{SolutionConsEqDM-OmegaDM-AlphaNeqZeta0-Gdm1}) of
Appendix~\ref{SectionHubbleParameter}, where all detailed calculations can
be found.

      \subsection{Cosmological data and
  $\chi^2$-functions \label{sec:cosm-data-chi2}}

To perform all numerical analysis we take, for the baryon and
radiation  components (photons and relativistic neutrinos), the values
of $\Omega_{b0} = 0.0458$\cite{WMAP7yKomatsu2011}, and $\Omega_{r0} =
0.0000766$, respectively, where the latter value is computed from the
expression\cite{WMAP5yKomatsu2009}
\begin{equation}
  \Omega_{\rm r0} = \Omega_{\gamma 0} (1+0.2271 N_{\rm eff}) \, .
\end{equation}
Here, $N_{\rm eff} = 3.04$ is the standard number of effective
neutrino species\cite{WMAP7yKomatsu2011,SDSS7yReid2010}, and
$\Omega_{\gamma 0} = 2.469 \times 10^{-5} h^{-2}$ corresponds to the
present-day photon density parameter for a temperature of $T_{\rm cmb}
= 2.725$ K\cite{WMAP7yKomatsu2011}, with $h$ the dimensionless Hubble
constant: $h \equiv H_0 /(100 \mathrm{km/s/Mpc})$. 

\subsubsection{Type Ia Supernovae}
The luminosity distance $d_L$ in a spatially flat FRW Universe is
defined as
\begin{equation}
  d_L(z, \gamma_{\rm de}, \alpha, \zeta_0) = \frac{c(1+z)}{H_0} \int_0^z
\frac{dz'}{E(z')}
\end{equation}
where $c$ corresponds to the speed of light in units of
$\mathrm{km/sec}$. The theoretical distance moduli $\mu^t$ for the
$k$~-~th supernova at a distance $z_k$ is given by
\begin{equation}
  \mu^t(z, \gamma_{\rm de}, \alpha, \zeta_0) = 5 \log \left[ \frac{d_L(z)
}{\rm Mpc} \right] +
  25 \, .
\end{equation}
Hence, the $\chi^2$-function for the SNe is defined as
\begin{equation}\label{Chi2FunctionSNe}
\chi^2_{\rm SNe}(\gamma_{\rm de}, \alpha, \zeta_0) \equiv \sum_{k=1}^n
\left( \frac{\mu^t(z_k, \gamma_{\rm de}, \alpha, \zeta_0)-
\mu_k}{\sigmạ_k} \right)^2 \, ,
\end{equation}
where $\mu_k$ is the observed distance moduli of the $k$-th supernova,
with a standard deviation of $\sigma_k$ in its measurement. 

For our case, $n= 580$, as we are using the type Ia supernovae (SNe
Ia) in the Union2.1 data set of the Supernova Cosmology Project (SCP),
which is composed of $580$ SNe Ia \cite{SNe-Union2.1:Suzuki2011}. We
have considered a \emph{flat} prior probability distribution
function to marginalize $H_0$ (i.e., it is not assumed any particular
value of $H_0$).

\subsubsection{Cosmic Microwave Background
  Radiation \label{sec:cosm-micr-backgr}}
We use the shift parameter $R$ reported in Table~9
of\cite{WMAP7yKomatsu2011}, that is defined as
\begin{equation}
  R = \frac{ H_0 \sqrt{\Omega_{\rm m0}}}{c} (1+z_*) D_A(z_*) \, ,
\end{equation}
where $\Omega_{\rm m0}$ is the present value of the density parameter
of the all pressureless matter in the Universe, i.e. $\Omega_{\rm m0}
= \Omega_{\rm b0} + \Omega_{\rm dm0}$, and $D_A$ is the proper angular
diameter distance given by
\begin{equation}
  D_A(z)= \frac{c}{(1+z)H_0} \int^z_0 \frac{dz'}{E(z')} \, ,
\end{equation}
in a spatially flat Universe. We then define a $\chi^2$-function as
\begin{equation}
  \label{Chi2FunctionRCMB}
  \chi^2_{\rm R-CMB}(\gamma_{\rm de}, \alpha, \zeta_0) \equiv \left(
\frac{R - R_{\rm obs}}{\sigma_{R}}
  \right)^2 \, ,
\end{equation}
where $R_{\rm obs} = 1.725$ is the \emph{observed} value of the shift
parameter, and $\sigma_{R}=0.018$ is its standard deviation
(cf. Table~9 of\cite{WMAP7yKomatsu2011}).

\begin{table*}[tp!]
  \centering
  \begin{tabular}{c c  c  c | c c | c c c c}
    \multicolumn{5}{c}{\textbf{SNe}}\\
    \hline \hline
    Model & $\gamma_{\rm de}$ & $\alpha$ & $\zeta_0$ & $\chi^2_{{\rm
        min}}$ & $\chi^2_{{\rm d.o.f.}}$ & DE & Energy Transfer & LSLT
    & CCD \\
    \hline
    I & $0^*$ & $-0.0132^{+0.22}_{-0.37}$   & $0.0017^{+0.097}_{-0.075}$
    & 562.223 & 0.972 & $\Lambda$ & DE $\leftarrow$ DM & \checkmark &
    \xmark \\
    II & $-0.0011^{+0.1}_{-0.11}$ & $-0.0086^{+0.1}_{-0.11}$ &  $0^*$ &
    562.224  & 0.972 & Phantom & DE $\leftarrow$ DM & \checkmark &
    \checkmark \\
    III & $-0.0040\pm{0.14}$ & $\alpha=\zeta_0$ &
    $-0.0026^{+0.035}_{-0.032}$ & 562.224 & 0.972 & Phantom & DE
    $\leftarrow$ DM & \xmark & \xmark \\
    IV & $-0.0052\pm{0.18}$ & $0^*$ & $-0.0037^{+0.055}_{-0.051}$ & 562.225
    & 0.972 & Phantom & None & \xmark & \xmark \\
    \hline
    \multicolumn{5}{c}{\textbf{$H(z)$}} \\
    \hline \hline
    I & $0^*$ & $-0.632^{+0.56}_{-1.17}$ &  $0.193^{+0.14}_{-0.15}$ &
    8.111 & 0.737 & $\Lambda$ & DE $\leftarrow$ DM & \checkmark &
    \xmark \\
    II & $-0.1976^{+0.13}_{-0.15}$ & $-0.0127^{+0.07}_{-0.09}$ & $0^*$ &
    8.046 & 0.731 & Phantom & DE $\leftarrow$ DM & \checkmark & \checkmark
    \\
    III & $-0.199^{+0.15}_{-0.16}$ & $\alpha=\zeta_0$ & $-0.0033 \pm 0.022$
    & 8.049 & 0.731 & Phantom & DE $\leftarrow$ DM & \xmark & \xmark
    \\
    IV & $-0.2006^{+0.16}_{-0.17}$ & $0^*$ &  $-0.00466^{+0.033}_{-0.031}$
    & 8.051 & 0.731 & Phantom & None & \xmark & \xmark \\
    \hline \\    
    \multicolumn{6}{c}{\textbf{SNe + CMB + BAO + $H(z)$}}\\
    \hline \hline
    I & $0^*$ & $0.0324^{+0.024}_{-0.025} $  & $-0.0085 \pm 0.005$
    & 572.766 & $0.965$ & $\Lambda$ & DE $\to$ DM & \xmark & \xmark \\
    II & $-0.0628^{+0.047}_{-0.049}$  & $-0.0112^{+0.012}_{-0.013}$
    & $0^*$ & 574.219 & 0.968 & Phantom & DE $\leftarrow$ DM &
    \checkmark & \checkmark\\
    III & $-0.0589^{+0.043}_{-0.045} $ & $\alpha = \zeta_0$ 
    & $-0.0023 \pm 0.0019$  & 573.618 & 0.967 & Phantom & DE
    $\leftarrow$ DM & \xmark & \xmark \\
    IV & $-0.0573^{+0.043}_{-0.044} $  & $0^*$ &  $-0.0028 \pm 0.002$
    & 573.522 & 0.967 & Phantom & None & \xmark & \xmark \\
    \hline
    V & $ (-6.04 \times 10^{-9}) \pm 0.05 $ & $0.028 \pm 0.03$ & $-0.008
    \pm 0.006$ & 571.199 & 0.965 & Phantom
    & DE $\to$ DM & \xmark & \xmark \\
    $\Lambda$CDM & $0^*$ & $0^*$  &  $0^*$  &  573.572 & 0.970 &
    $\Lambda$ & None & \checkmark & \checkmark \\
    \hline
  \end{tabular}
  \caption{\label{TableBestEstimated} Marginal best estimated values of the
    parameters $(\gamma_{\rm  de}, \alpha, \zeta_0)$ for the different
    models discussed in the text; notice that the DM barotropic index
    is that of a dust fluid for all cases, $\gamma_{\rm DE} = 1$. The
    asterisk superscript indicates the cases when the zero value of
    one of the parameter was assumed a priori. The top (middle) table
    only considers SNe ($H(z)$) observations, whereas the bottom table
    correspond to the use of the combined SNe + CMB + BAO + $H(z)$
    data sets together, see Sec.~\ref{SectionCosmologicalObservations}. The
    fourth and fifth columns correspond to the minimum value of the
    $\chi^2$ function, $\chi^2_{\rm min}$, and the $\chi^2$ by degrees of
    freedom, $\chi^2_{\rm d.o.f.}$, respectively. The latter is
    defined as $\chi^2_{\rm d.o.f.}= \chi^2_{\rm min} / (n-p)$, where
    $n$ is the number of data and $p$ the number of free
    parameters. The next-to-last row (Model V) corresponds to best estimates
of the three parameters $(\gamma_{\rm  de}, \alpha, \zeta_0)$ computed
simultaneously. $H_0$ was marginalized assuming a flat prior distribution.
 The last row, with $(\gamma_{\rm de}=0, \alpha=0,
    \zeta_0=0)$ corresponds to the value that we obtain for the
    $\Lambda$CDM model, using the same procedure and data sets, in
    order to compare our results. According to the value $\chi^2_{\rm
      d.o.f.}$, we find that all our cases fit the data sets as well
    as $\Lambda$CDM does. Last columns indicate the type of DE, the
    energy transfer direction, the consistency with the Local Second Law of
    Thermodynamics (LSLT, Eq.~(\ref{entropy_condition})), and with a
    complete cosmological dynamics (CCD) as discussed in
    Sec.~\ref{wellbehavecondition}. See Figs.~\ref{PlotGroupI} to
    \ref{PlotGroupIV} for their corresponding confidence intervals,
    and the text for more details.}      
\end{table*}

\subsubsection{Baryon Acoustic Oscillations \label{sec:bary-acoust-oscill}}
We use the baryon acoustic oscillation (BAO) data from the SDSS
7-years release \cite{Percival:2009xn}, expressed in terms of the distance
ratio $d_z$ at $z=0.275$, which is defined as
\begin{equation}
  d_{0.275} \equiv \frac{r_s(z_d)}{D_V(0.275)} \, ,
\end{equation}
where $r_s(z)$ corresponds to the comoving sound horizon given by
\begin{equation}
  r_s(z)=\frac{c}{\sqrt{3}} \int_0^{1/(1+z)} \frac{da}{a^2 H(a)
    \sqrt{1+(3 \Omega_{\rm b0} / 4 \Omega_{\gamma 0})a}} \, .
\end{equation}
As mentioned above, we take the following actual values of the density
parameters: $\Omega_{\gamma 0} = 2.469 \times 10^{-5}
h^{-2}$ for photons, and $\Omega_{b0} = 0.02255 h^{-2}$ for
baryons\cite{WMAP7yKomatsu2011}.
And $z_d$ is the redshift at the baryon drag epoch computed from the
fitting formula\cite{Eisenstein:1997ik}
\begin{subequations}
\begin{eqnarray}
  z_d &=& 1291 \frac{(\Omega_{\rm m0} h^2)^{0.251}}{1+0.659(\Omega_{\rm m0}
    h^2)^{0.828}} \left[ 1 + b_1 (\Omega_{\rm m0} h^2)^{b_2} \right]
  \, , \\
  b_1 &=& 0.313 (\Omega_{\rm m0} h^2)^{-0.419} \left[1 + 0.607 (\Omega_{\rm
      m0} h^2)^{0.674} \right] \, , \\
  b_2 &=& 0.238 (\Omega_{\rm m0} h^2)^{0.223} \, .
\end{eqnarray}
\end{subequations}
For a flat Universe, $D_V(z)$ is defined as
\begin{equation}
  D_V(z) = c \left[ \left( \int_0^z \frac{dz'}{H(z')} \right)^2
    \frac{z}{H(z)} \right]^{1/3} \, .
\end{equation}
It contains the information of the visual distortion of a spherical
object due the non-Euclidianity of the FRW spacetime. Parameter
$d_{0.275}$ contains the information of the other two pivots,
$d_{0.2}$, and $d_{0.35}$, that are usually used by other authors,
with a precision of $0.04\%$ \cite{Percival:2009xn}.

The $\chi^2$ function for BAO is then given by
\begin{equation}
  \label{Chi2FunctionBAO}
  \chi^2_{\rm BAO}(\gamma_{\rm de}, \alpha, \zeta_0) \equiv \left(
    \frac{d_{0.275} - d_{0.275}^{\rm obs}}{\sigma_{d}} \right)^2 \, ,
\end{equation}
where $d_{0.275}^{\rm obs} = 0.139$ is the observed value, and
$\sigma_{d} = 0.0037$ is its standard deviation\cite{Percival:2009xn}.

\subsubsection{Hubble expansion rate \label{SectionHz}}
For the Hubble parameter we use the $13$ available data; $11$ data
come from Table~2 in Stern et al. (2010)\cite{Hz-Stern:2009}, and
other 2 from Gaztanaga et al. (2010)\cite{Hz-Gaztanaga:2008}:
$H(z=0.24)=79.69 \pm 2.32$ and $H(z=0.43)= 86.45 \pm 3.27$
km/s/Mpc. For the present value of the Hubble parameter, we take the
value reported by Riess et al (2011)\cite{HzToday-Riess:2011}: $H(z=0)
\equiv H_0 = 73.8 \pm 2.4$ km/s/Mpc. The $\chi^2$ function is
\begin{equation}
  \label{Chi2FunctionHz}
  \chi^2_{\rm H}(\gamma_{\rm de}, \alpha, \zeta_0) = \sum_i^{13} \left(
\frac{H(z_i) - H_i^{\rm
        obs}}{\sigma_{H}} \right)^2 \, ,
\end{equation}
where $H(z_i)$ is the theoretical value predicted by the model, and
$H_i^{\rm obs}$ is the observed value with a standard deviation
$\sigma_{H}$.

\subsection{Observational constraints \label{sec:observ-constr-}}
Finally, with each of the $\chi^2$-functions defined above we
construct the total $\chi^2$-function given by
\begin{equation}\label{Chi2FunctionTotal}
\chi^2 = \chi^2_{\rm SNe} + \chi^2_{\rm CMB} + \chi^2_{\rm BAO} +
\chi^2_{\rm H}.
\end{equation}
We minimize this function with respect to the parameters $(\gamma_{\rm de},
\alpha, \zeta_0)$ to compute their best estimated values and confidence
intervals.

\begin{figure*}[ht!]
\includegraphics[width=7cm]{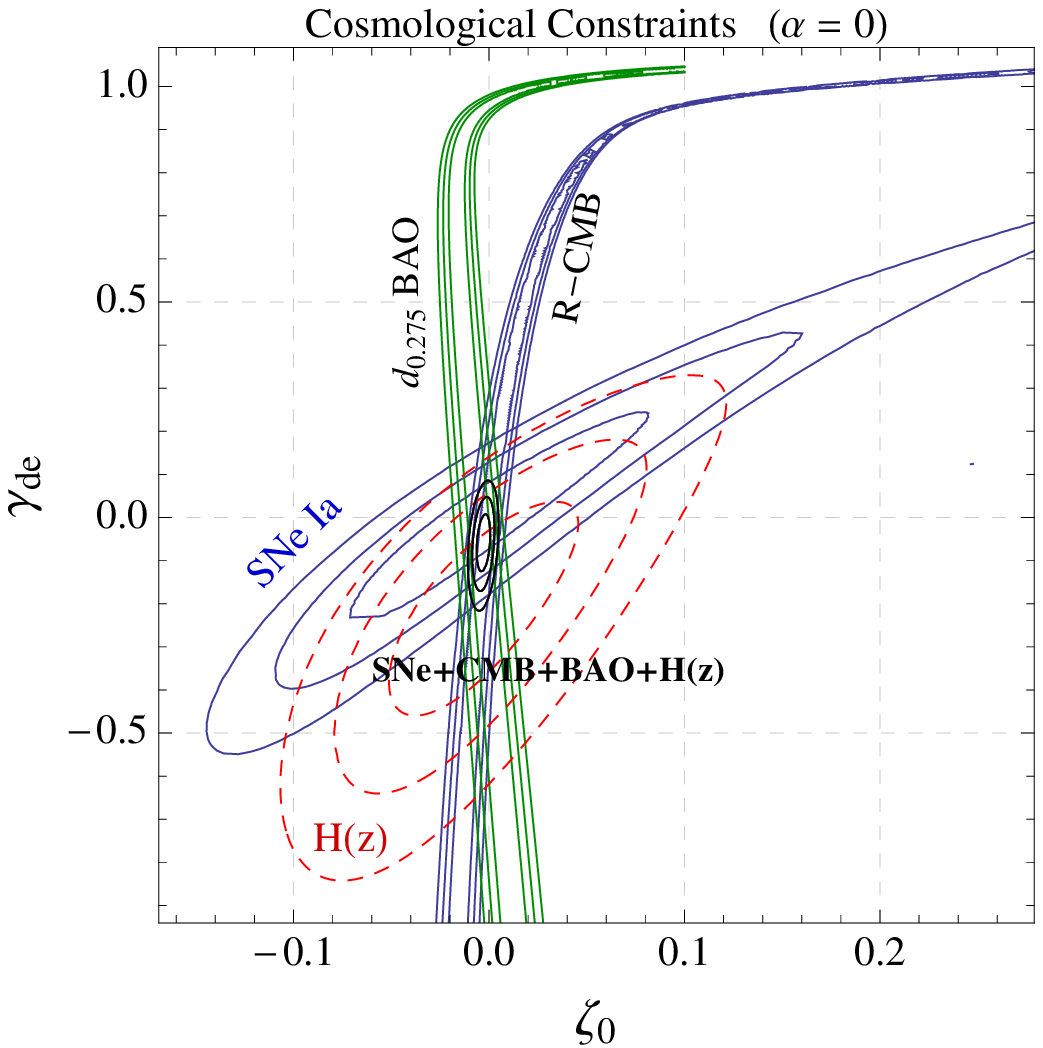}%
\includegraphics[width=7cm]{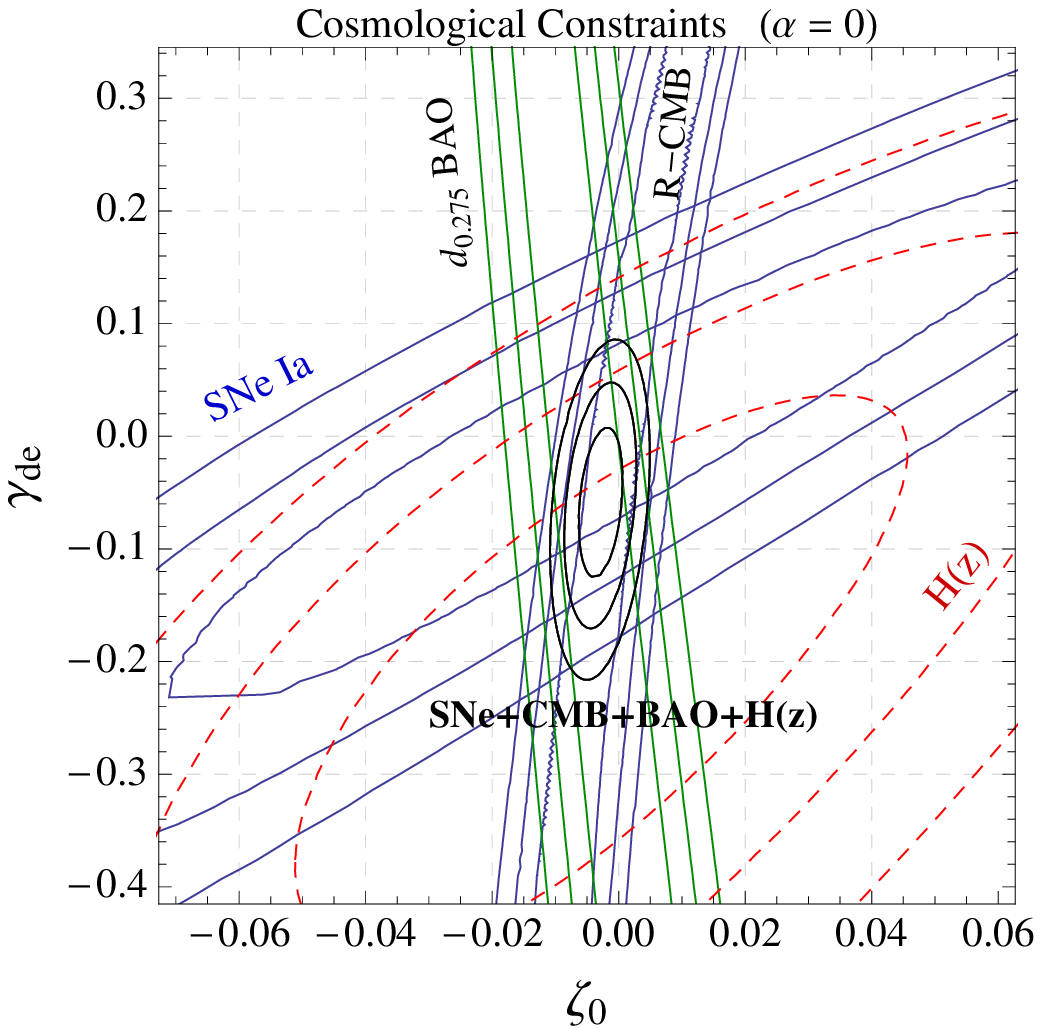}%
\caption{\label{PlotGroupIV} 
Confidence intervals (CI) for Model IV: $(\zeta_0, \gamma_{\rm de})$
as free parameters, and $\alpha =0$. The CI correspond to $68.3 \%$,
$95.4 \%$ and $99.7 \%$ of confidence level. We notice from this
figure that, with $99.7 \%$, and using the combined SNe + CMB + BAO +
$H(z)$ data sets together,  the values lie on the regions $  -0.011<
\zeta_0 < 0.005$ and $ -0.22 <  \gamma_{\rm de} <  0.085$,  when
$(\zeta_0, \gamma_{\rm de})$ are constrained simultaneously. The
marginal best estimated values for each parameter individually are
$\zeta_{\rm 0} = -0.0028 \pm 0.002$, and $\gamma_{\rm de}=
-0.0573^{+0.043}_{-0.044})$, where the errors are given to $68.3 \%$
of confidence, see also Table~\ref{TableBestEstimated}. (Right)
\textit{Zoom in}  of the CI around the best estimated values. The bulk
viscosity is constrained again to small values and mainly in the
negative region, with almost $68\%$ of probability ($1\sigma$), that
is in tension with the LSLT and with the requirement of a well behaved
cosmology from the dynamical system analysis. So, Model IV is ruled
out with almost $1\sigma$. On the other hand, we find that
values of $\gamma_{\rm DE} < 0$ are preferred by the observations with
a $68\%$ of probability, corresponding to a \textit{phantom} dark
energy.} 
\end{figure*}

There are four special cases we will discuss here, whose parameters
are described and estimated in Table~\ref{TableBestEstimated}, and in
the confidence intervals (CI) in Figs.~\ref{PlotGroupI},
\ref{PlotGroupII}, \ref{PlotGroupIII}, and~\ref{PlotGroupIV}.

Some general comments are in turn before the detailed explanation of
the different models. First, we have noticed, for the
quantities reported in Table~\ref{TableBestEstimated}, that there is a
qualitative change in the models if only the low-redshift data sets are
taken into account; in our case, these data sets are those of the
supernovae (SNe) and the Hubble parameter ($H(z)$). Such change is
particularly acute in the case of the bulk viscosity $\zeta_0$: it is
consistently positive definite whenever the interaction parameter
$\alpha$ is set free, like in Model I. If $\alpha$ is fixed to be
equal to $\zeta_0$, or to have a null value, then the bulk viscosity is
negative definite, like in Models III and IV.

But right the opposite happens if high-redshift measurements are
included in the analysis: all models consistently point out to a
negative value of the bulk viscosity whenever it is freely fitted,
like in Models I, III, and IV. This means, actually, that all models
with bulk viscosity as a free parameter are at variance with the
LSLT when they are fitted to the sample set of cosmological observations.

Our second general comment is that none of the models is consistent
with our so-called statement of complete cosmological dynamics
presented and discussed in Sec.~\ref{wellbehavecondition}. The main
reason being that we cannot recover an appropriate RDE at early
times. One must notice, though, that our data sets cannot cover high
enough redshifts in order to properly sample the early RDE of the
Universe, but it is nonetheless significant that the estimated values
of the free parameters already indicate a non-recovery of an RDE. Such
a difficulty was already observed in models with bulk
viscosity\cite{AvelinoEtal2013,Visco-VeltenSchwarz-2012} , but
it has not been sufficiently remarked in models with a DM-DE
interaction\cite{Chimento:2007yt,Quartin2008,Caldera-Cabral2009}.

One last comment regards that of the nature of DE in all models: we
have consistently found that phantom
DE\cite{Caldwell:1999ew,*Scherrer:2004eq} is slightly favored by all
data sets whenever the DE EOS is freely varied, and that in most of
the models an energy transfer from DM to DE is preferred.


\subsubsection{Model I \label{sec:model-i-}}
Model I corresponds to $\gamma_{\rm DE} = 0$, whereas $\alpha$ and
$\zeta_0$ are free parameters; that is, this case corresponds to a
DE-DM interacting model, in which DM is a dust fluid with bulk
viscosity, and DE is a cosmological constant, see for
instance\cite{Visco-Kremer2012,Rashid:2009td,Adabi:2011eh} and
references therein for similar models. 

\begin{figure*}[htp!]
\includegraphics[width=7cm]{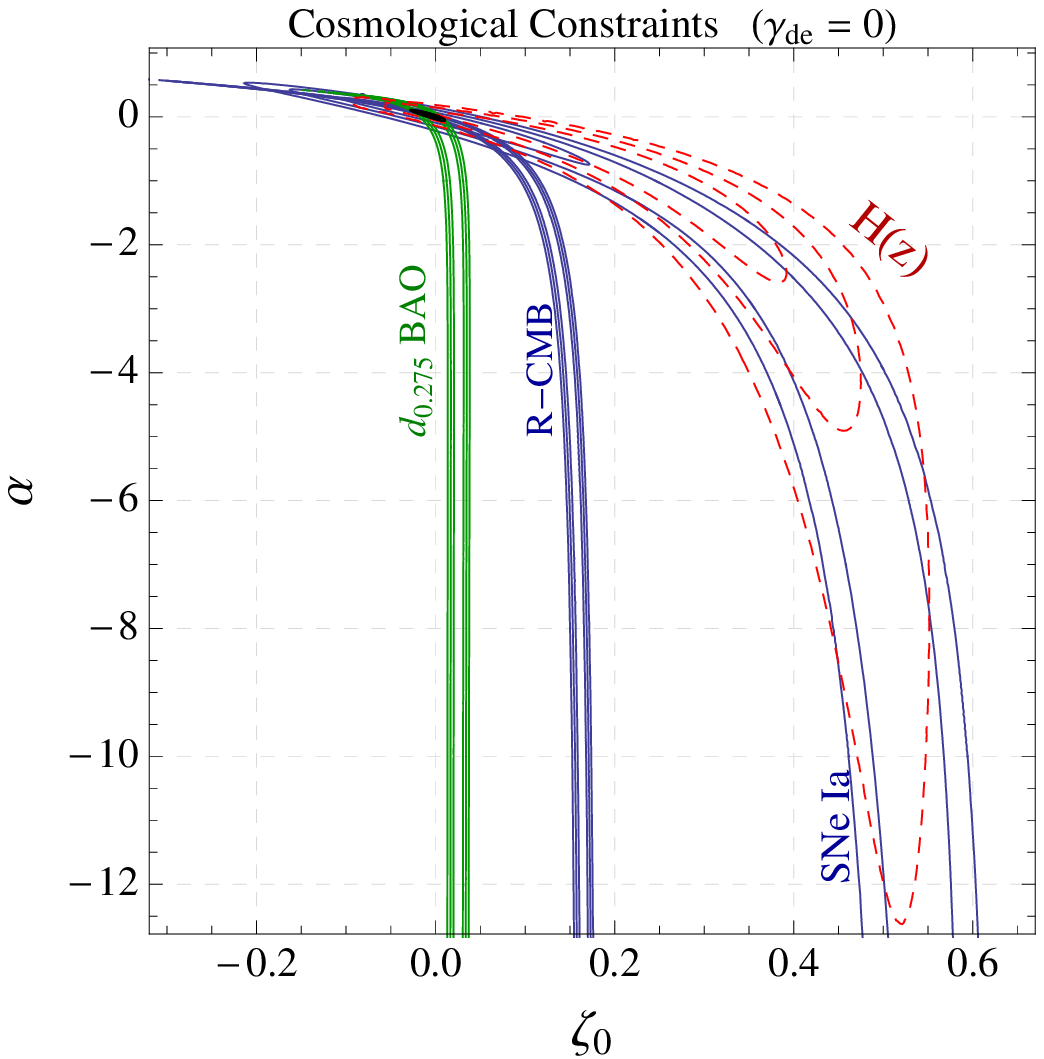}%
\includegraphics[width=7cm]{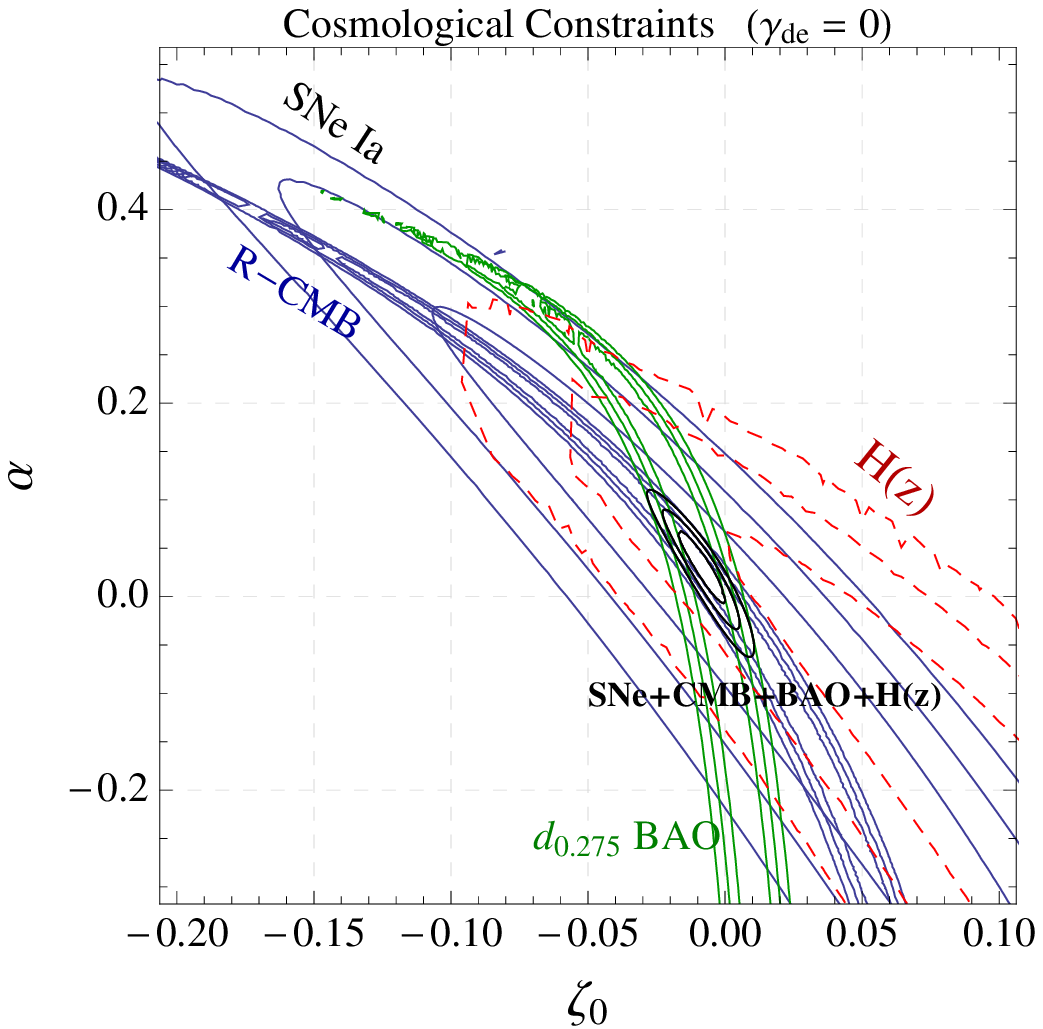}%
\caption{\label{PlotGroupI} Confidence intervals (CI) for Model I:
  $(\zeta_{\rm 0}, \alpha)$ as free parameters, and $\gamma_{\rm DE} =
  0$. The CI shown  correspond to $68.3 \%$, $95.4 \%$ and $99.7 \%$
  of confidence level. We notice from  this figure that, with $99.7
  \%$ of confidence, and using the combined SNe + CMB + BAO + $H(z)$
  data sets together, the values lie on the regions $ -0.029 <
  \zeta_{\rm 0} < 0.011$ and $ -0.062 <  \alpha < 0.11$, when
  $(\zeta_0, \alpha)$ are constrained simultaneously. The marginal
  best estimated values for each  parameter individually are
  $\zeta_{\rm 0} = -0.0085 \pm 0.005$, and
  $\alpha=0.0324^{+0.024}_{-0.025}$, where the errors are given to
  $68.3\%$ of confidence, see also
  Table~\ref{TableBestEstimated}. (Right) \textit{Zoom in} of the CI
  around the best estimated values. The bulk viscosity is constrained
  to small negative values; however, $\zeta_0 < 0$ is forbidden by the
  LSLT, and then Model I is ruled out with a $68 \%$ of probability
  ($1\sigma$). We notice that there is a slight preference for small
  but positive values of $\alpha$. Finally, the CI for $(\zeta_{\rm
    0}, \alpha)$ parameters lie almost completely in a region that is
  not consistent with a well behaved cosmology defined by the
  dynamical system analysis, and then Model I must be considered to be
  ruled out.}
\end{figure*}

According to the values presented in Table~\ref{TableBestEstimated}
and in Fig.~\ref{PlotGroupI}, the bulk viscosity is positive for
low-redshift data sets, but it takes small negative values when the
full data set is considered. However, we must recall that $\zeta_0 <
0$ is forbidden by the LSLT, see Eq.~(\ref{entropy_condition}), and
because of this Model I would then be ruled out with at least $68 \%$ of
probability ($1\sigma$). We notice that there is a slight preference
for small but positive values of $\alpha$, i.e, from the figure
\ref{PlotGroupI} we see that the $68 \%$ contour region lies in the positive
region for $\alpha$. Finally, the CI for
$(\zeta_{\rm 0}, \alpha)$ parameters lie almost completely in a region
that is not consistent with a well behaved cosmology defined by the
dynamical system analysis in Sec.~\ref{wellbehavecondition}, because
none of the critical points $P_{1a}$, $P_{2a}$ or $P_{1c}$, see Table~\ref{tab2},
is a suitable point for a RDE; this fact then adds for the
ruling out of this model.

\subsubsection{Model II \label{sec:model-ii-}}
Model II corresponds to $\zeta_0 =0$, whereas $\gamma_{\rm DE}$ and
$\alpha$ are free parameters; that is, it corresponds to a purely
DM-DE interacting model, see for instance\cite{Quartin2008,
  Caldera-Cabral2009,*Caldera-Cabral2010,Avelino2013} and references
therein. By definition, this model is in agreement with the LSLT.

Both parameters $(\gamma_{\rm de}, \alpha)$ are close to zero, but
\textit{phantom} DE is slightly favored, $(\gamma_{\rm DE} <0)$ at
about $68.3 \%$ (1$\sigma$), as also is $\alpha < 0$, which
corresponds to energy transfer from DM to DE. In both region, as shown
in Table~\ref{tab2}, the model describes a \emph{complete cosmological
  dynamics} since is possible to choose initial conditions that lead
orbits to connect $P_{1a}$$\rightarrow$
$P_{2b}$$\rightarrow$$P_{3a}$.

\begin{figure*}[ht!]
\includegraphics[width=7cm]{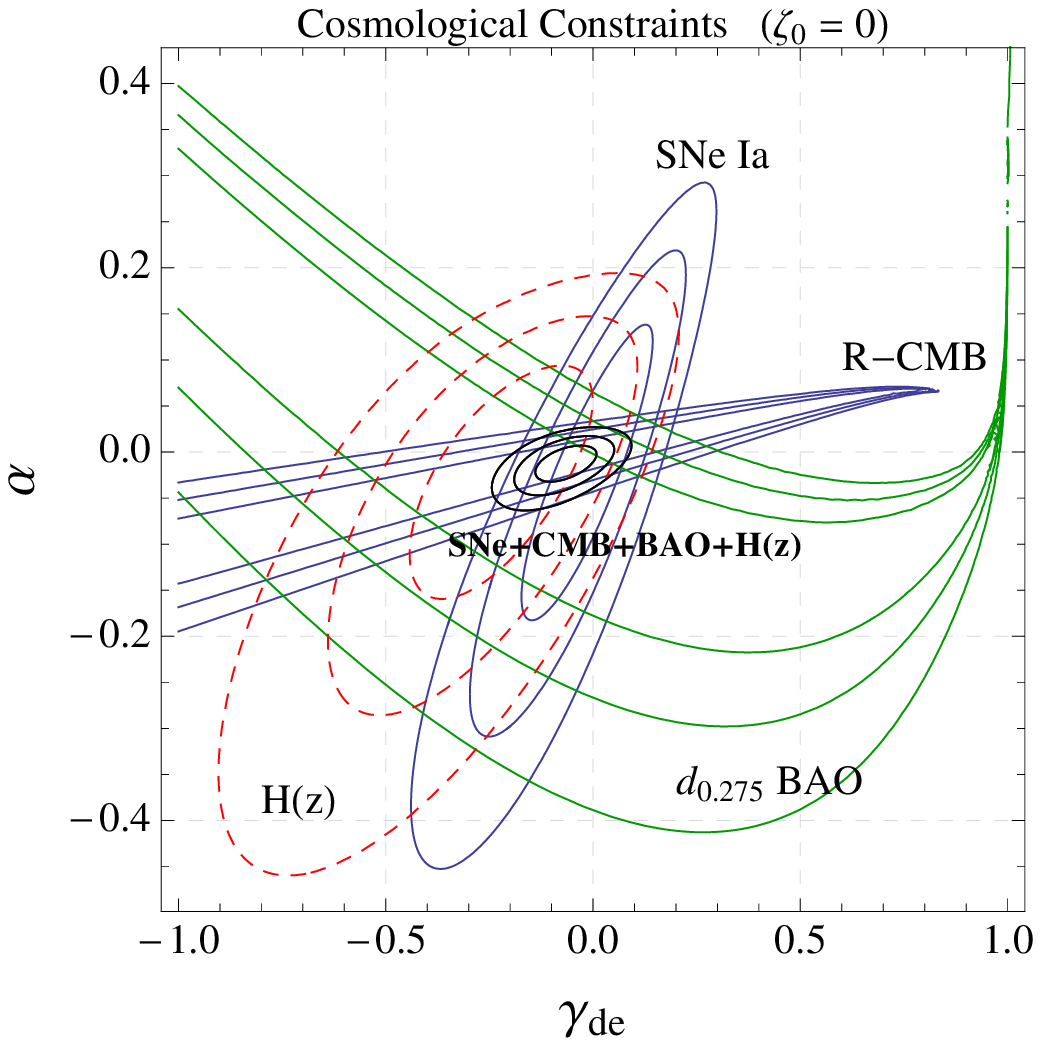}%
\includegraphics[width=7cm]{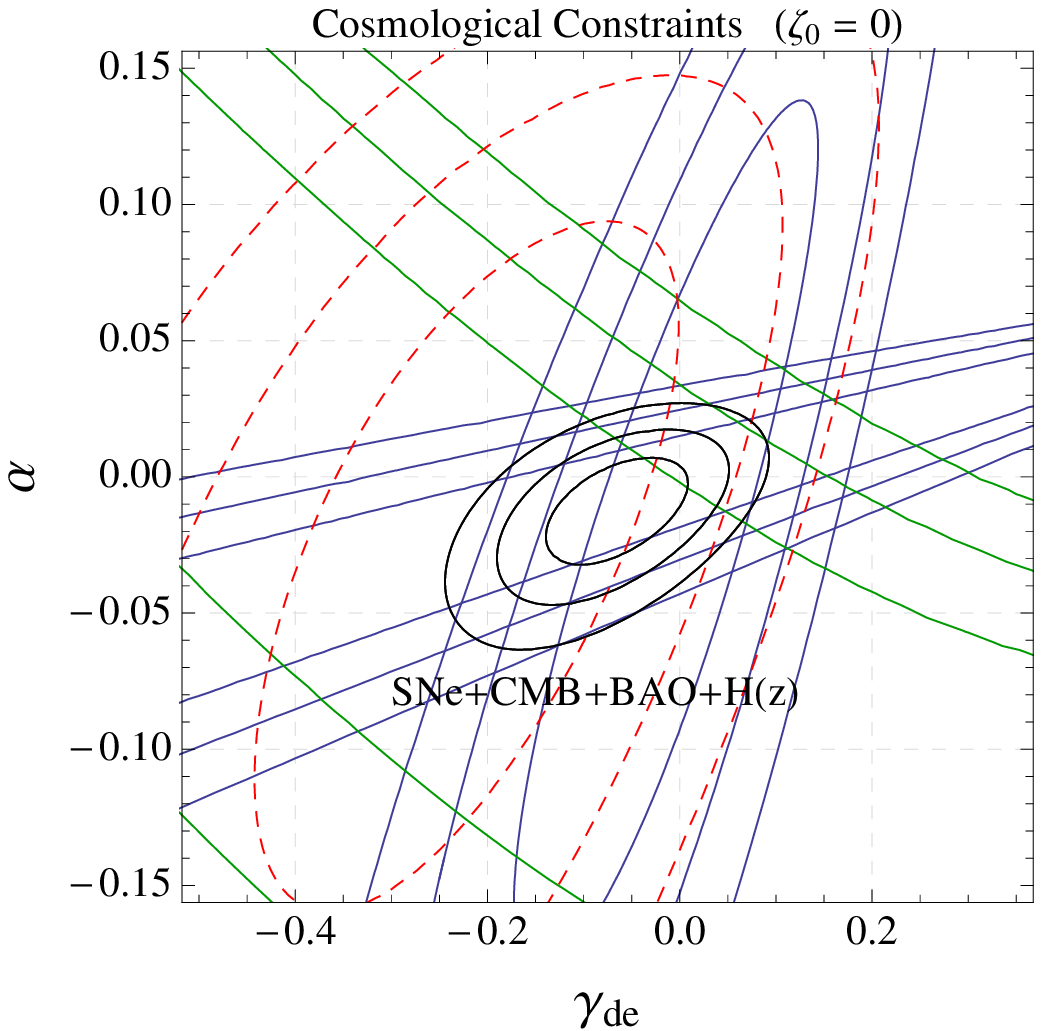}%
\caption{\label{PlotGroupII} 
Confidence intervals for Model II (DM-DE interacting model
\textit{without} bulk viscosity): $(\gamma_{\rm de}, \alpha)$ as free
parameters, and $\zeta_{\rm 0} =0$. The CI correspond to $68.3 \%$,
$95.4 \%$ and $99.7 \%$ of confidence level. We notice that, with
$99.7 \%$, and using the combined SNe + CMB + BAO + $H(z)$ data sets
together, the values lie on the regions $  -0.25<  \gamma_{\rm de} <
0.10 $ and $-0.065  < \alpha< 0.028$ , when  $(\gamma_{\rm de},
\alpha)$ are constrained simultaneously. The marginal best estimated
values for each parameter individually are $\gamma_{\rm de} =
-0.0628^{+0.047}_{-0.049}$, and $\alpha=-0.0112^{+0.012}_{-0.013})$, where
the errors are given to $68.3 \%$ of confidence, see also
Table~\ref{TableBestEstimated}. (Right) \textit{Zoom in} of CI around
the best estimated values. Both parameters $(\gamma_{\rm de}, \alpha)$
are close to zero, but \textit{phantom} DE is slightly favored,
$\gamma_{\rm DE} <0$, at about $68.3 \%$ (1$\sigma$), as also is
$\alpha < 0$, which corresponds to energy transfer from DM to
DE.}
\end{figure*}

\subsubsection{Model III \label{sec:model-iii-}}
Model III corresponds to $\alpha = \zeta_0$, whereas $\gamma_{\rm DE}$
and $\zeta_0$ are free parameters; that is, this case corresponds to
an \textit{interacting} bulk viscous DM-DE model, where the
interacting parameter is directly proportional to the bulk
viscosity. A related model was studied by Kremer and
Sobreiro\cite{Visco-Kremer2012} where they assume $\alpha =
-\zeta_0$.

\begin{figure*}[ht!]
\includegraphics[width=7cm]{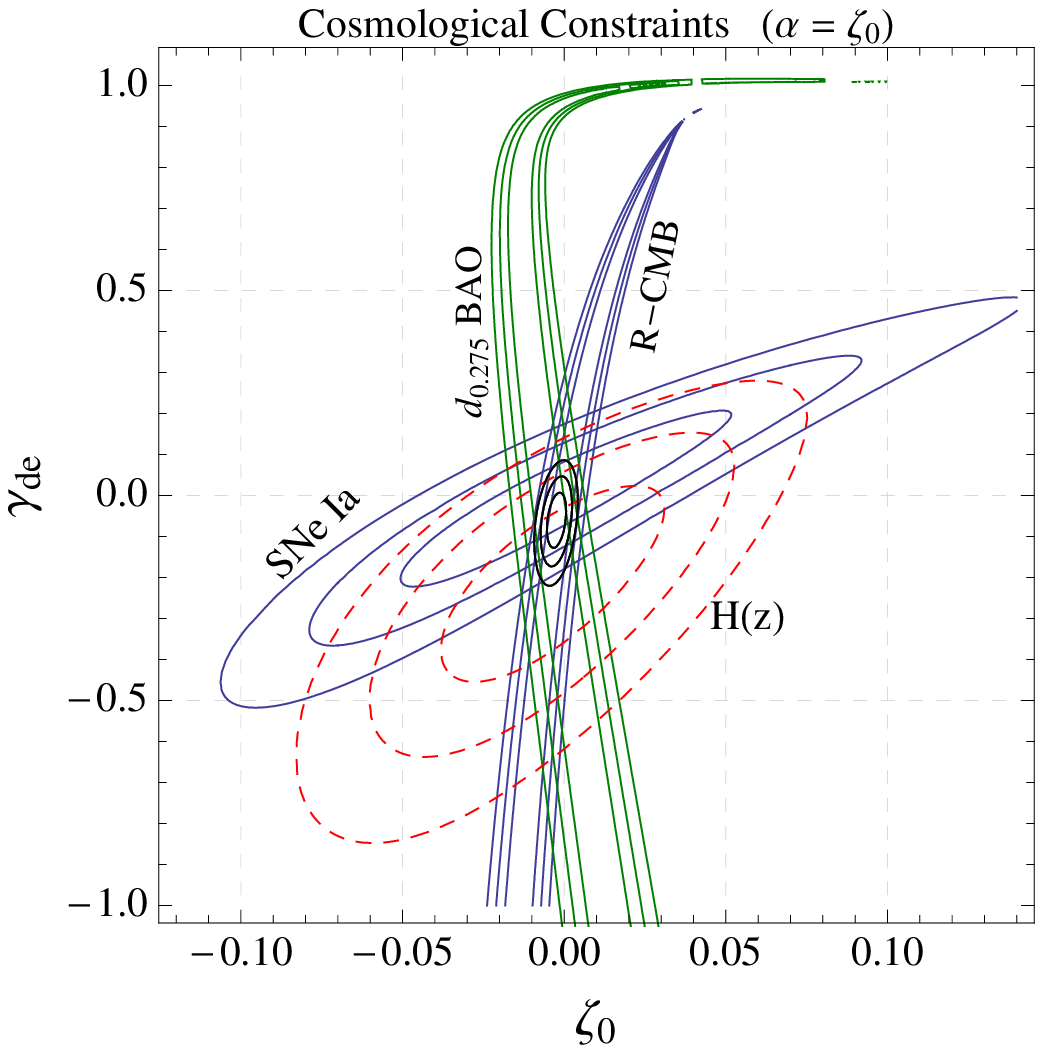}%
\includegraphics[width=7cm]{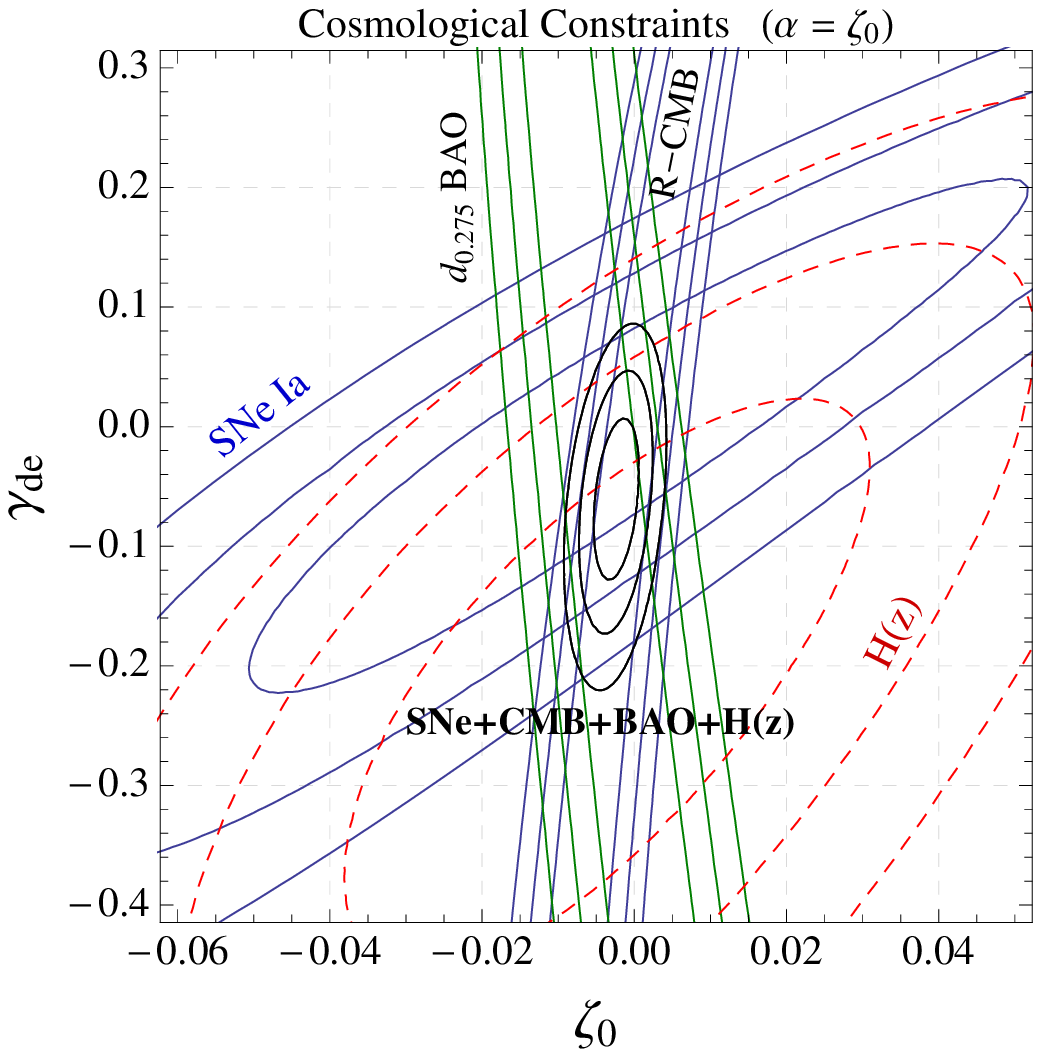}%
\caption{\label{PlotGroupIII} 
  Confidence intervals (CI) for Model III: $(\zeta_{\rm 0}, \gamma_{\rm
    de})$ as free parameters, and $\alpha= \zeta_{\rm 0}$. The CI shown 
  correspond to $68.3 \%$, $95.4 \%$ and $99.7 \%$ of confidence
  level. We notice that, with $99.7 \%$, and using the combined SNe +
  CMB + BAO + $H(z)$ data sets together, the values lie in the regions
  $ -0.011 < \zeta_{\rm 0} < 0.005$ and $ -0.22 <  \gamma_{\rm de} <
  0.09$, when $(\zeta_{\rm 0}, \gamma_{\rm de})$ are constrained
  simultaneously. The marginal best estimated values for each
  parameter individually are $(\zeta_{\rm 0}=-0.0023  \pm 0.0019 ,
  \gamma_{\rm de}= -0.0589 ^{+0.043}_{-0.045})$, where the errors are
  given to $68.3 \%$ of confidence, see also
  Table~\ref{TableBestEstimated}. (Right) \textit{Zoom in}  of
  the CI around the best estimated values. We find interesting that
  the CI's are almost identical to the case $\alpha = 0$ (see
  Fig.~\ref{PlotGroupIV}), suggesting that the value of the bulk
  viscosity and the nature of the DE in this model is insensitive to
  the assumption of the interaction. However, this model must be
  considered to be ruled out because is at variance with the LSLT
  ($\zeta_{0}<0$) and is not consistent with a \emph{complete
    cosmological dynamics}.}
\end{figure*}

We find interesting that the CI's presented in Fig.~\ref{PlotGroupIII}
are almost identical to those of Model II (for $\alpha = 0$), see
Sec.~\ref{sec:model-ii-} and Fig.~\ref{PlotGroupII}, suggesting that
the value of the bulk viscosity and the nature of the DE component in
this model is insensitive to the assumption of the DM-DE
interaction. This model is not compatible with a RDE, as Table
~\ref{tab2} shown. In addition, is not possible to recover a true MDE
since only $P_{2a}$ is fulfilled, namely a pure DM domination with a
null contribution of baryonic matter. Regardless of the initial
conditions, $P_{3a}$ is the only possible late time attractor of Model
III, but, as Table~\ref{TableBestEstimated} shows, observations favor
negative values for $\alpha$ (and hence for $\zeta_0$) and, for those
negative values $P_{3a}$ does not belong to the phase space
(\ref{eq:space}) of the model. Thus, the model is ruled out because it
is neither compatible with a \emph{complete cosmological dynamics} nor
with the LSLT.

\subsubsection{Model IV \label{sec:model-iv-}}
Model IV corresponds to $\alpha = 0$, whereas $\gamma_{\rm DE}$ and
$\zeta_0$ are free parameters; that is, it corresponds to a
\textit{non}-interacting DM-DE model, in which DM has bulk viscosity.
See for instance \cite{Kremer:2002hz,Cataldo:2005qh} and references
therein.
 
From the CI of the joint SNe + CMB + BAO + $H(z)$ datasets we find that the
bulk viscosity is constrained again to small values and mainly in the
negative region, with almost $68\%$ of probability ($1\sigma$), that is in
tension with the LSLT and our statement of a complete cosmological
dynamics. On the other hand, we find that values of $\gamma_{\rm de} <
0$ are preferred by the observations with at least $68\%$ of probability,
corresponding to \textit{phantom} DE.

\subsubsection{Model V and $\Lambda$CDM \label{sec:model-v-lambdacdm}}
Model V corresponds to the case in which all parameters are freely
varied simultaneously, and as such is our most general case. As in
previous cases, we find again that phantom DE is slightly preferred,
as is also the energy transfer from DE to DM. However, the final
output is not compatible with the LSLT nor with a \emph{complete
  cosmological dynamics}; the latter mainly because a proper RDE
cannot be recover at early time.

Just for comparison, we have also fitted the $\Lambda$CDM model to the
same data and using the procedure; notice that this model is also our
null-hypothesis case, as
it is recovered if all parameters are given null values. Interestingly
enough, the good of fitness of our models is as good as that of
$\Lambda$CDM, a fact that points out that the used data sets are not
powerful enough to differentiate the models; this is why we had to
consider other constraints from the theoretical point of view, like
that of the LSLT in Eq.~(\ref{entropy_condition}), and the complete
cosmological dynamics reviewed in Sec.~\ref{sec:cosm-evol-from}.

\section{Discussion and Conclusions}\label{SectionConclusions}
In the present work we studied, in general terms, a cosmological model
that includes DM with bulk viscosity, an interaction term between DM
and DE, and a free barotropic equation of state $p = (\gamma_{\rm de}
- 1) \rho_{\rm de}$. The dissipation in the DM component was
characterized by a bulk viscosity $\zeta$ directly proportional to the
expansion rate of the Universe, i.e.,  $\zeta = H \zeta_0 /(8 \pi G)$, where
$\zeta_0$ is a dimensionless constant. Another important assumption
was that, except the DM, all matter components are represented by perfect
fluids, which in itself constraints the type of DE that are affected by
our analysis.

First of all, we performed a detailed dynamical system analysis of the
model in order to investigate its asymptotic evolution and
behavior. In addition, we demanded that our model must follow what we
called a \emph{complete cosmological dynamics}, namely: the existence
of a viable RDE and MDE prior to a late-time acceleration stage; these
three different eras have to be present in any model of physical
interest. The imposition of this requirement rules out any model with
a bulk viscosity in the DM sector. This results from the fact that the
bulk viscosity needs to be negative definite in order to have standard
RDE and MDE, but that is not possible if we are to believe in the
LSLT. However, a negative definite bulk viscosity is compatible with
the speed up of the Universe at low redshifts, which actually was one
of the appealing aspects of these type of models.

For purposes of illustration, we have applied our general results to
the specific interaction function: $Q= 3 \alpha \rho_{\rm de} H$,
where the parameter $\alpha$ quantifies the strength and direction of
the DM-DE interaction. As said before, we found that the bulk
viscosity parameter was the troublesome one, and that we could
accommodate a complete cosmological dynamics as long as $\zeta_0 =0$.

Also, we tested the model using cosmological observations to estimate
the free parameters and set constraints on them. The three parameters
($\gamma_{\rm de}, \alpha, \zeta_0$) allowed us to have a very rich
diversity of possible models to study, from purely interacting models
($\zeta_0 =0$), to purely viscous models ($\alpha =0$), and even the
case of $\Lambda$CDM, which then acted as our null hypothesis
($\zeta_0 =0 = \alpha$).


Whenever we tested a model with a non-null bulk viscosity, we found
that a {\it negative} value of it was preferred, with at least
1$\sigma$ of confidence level. This result is a drawback of the model,
given that it is in tension with the LSLT that reads $\zeta_0 >0 $ for
our model. It should be said, though, that such a result was obtained
if high-redshift data was included in the analysis. Actually,
low-redshift data seems to favor a positive definite value of the bulk
viscosity, but that would have lead us to wrong conclusions about the
viability of the model. As for the interaction parameter $\alpha$, we
found that in general the data favor a negative value, indicating an
energy transfer from the DM to DE. 


On the other hand, it is interesting to notice that using the
cosmological observations we consistently found \textit{negative}
values of the barotropic index $\gamma_{\rm de}$, suggesting a phantom
nature for the DE fluid that is in agreement with recent
results\cite{Planck}, even though such a setup is troublesome from the
theoretical point of view (like in the violation of the null energy
condition $\rho + p \geq 0$).

We computed also the $\chi^{2}_{\rm d.o.f.}$ of all the models, and
found that the goodness-of-fit to the data were equally good for all
of them. This fact seems to indicate that the inclusion of new free
parameters did not significantly improve the viability of the models,
nor did it help to distinguish them from the null hypothesis
represented by the concordance $\Lambda$CDM model.


\appendix*

\section{The Hubble parameter \label{SectionHubbleParameter}}
Here, we give details about the calculation of the Hubble parameter,
see~Eq.~(\ref{Chi2FunctionSNe}), that is required in
Sec.~\ref{SectionCosmologicalObservations} to compute the
observational constraints. 

The exact solutions of the conservation
equations~(\ref{ConsEqRadiation}), and~(\ref{ConsEqBaryon}), are,
respectively,
\begin{equation}
  \rho_{\rm r}(a) = \rho_{\rm r0}/a^4 \, , \quad \rho_{\rm b}(a) =
  \rho_{\rm b0}/a^3 \, , 
\end{equation}
where $a$ is the scale factor, and the subscript zero labels the
present values of the energy densities. If we take the interaction
term $Q=3H \alpha \rho_{\rm de}$, the conservation
equations~(\ref{EqConservationEffective4}),
and~(\ref{EqConservationEffective4a}) can be rewritten as
\begin{equation}
  \dot{\rho}_{\rm dm} + 3H\gamma^e_{\rm dm} \rho_{\rm dm} = 0 \,
, \quad \dot{\rho}_{\rm de} + 3H\gamma^e_{\rm de} \rho_{\rm de} = 0 \,
, \label{ConsEqDE}
\end{equation}
where we have defined the effective barotropic indexes
\begin{eqnarray}
  \gamma^e_{\rm dm} &=& \gamma_{\rm dm} + (\gamma_{\rm
    de} - \gamma^e_{\rm de}) \frac{\rho_{\rm de}}{\rho_{\rm dm}} -
  \frac{3H \zeta}{\rho_{\rm dm}} \,
  , \label{EqGralIndicesBarotropicos} \\
  \gamma^e_{\rm de} &=& \gamma_{\rm de} + \alpha \, . \label{GammaFr}
\end{eqnarray}

As all parameters are constant, we can integrate the equation of
motion for the DE energy density, see Eq.~(\ref{ConsEqDE}) and obtain
\begin{equation}\label{DensityDE2}
\rho_{\rm de}(a) = \rho_{\rm de0} a^{-3(\gamma_{\rm de} +\alpha)} \, .
\end{equation}
Hence, the barotropic index of DM, see
Eq.~(\ref{EqGralIndicesBarotropicos}), can be written as
\begin{equation}\label{Gammas2}
  \gamma^e_{\rm dm} = \gamma_{\rm dm} - \frac{1}{\rho_{\rm dm}}
  \left(\alpha \rho_{\rm de}+ \zeta_0\frac{3H^2}{8 \pi G}
  \right) \, .
\end{equation}
With the help of the Friedmann constraint~(\ref{ConstrainFriedmann}),
and the exact solutions of the energy densities, Eq.~(\ref{Gammas2})
can be finally rewritten as
\begin{multline}\label{Gammas4}
\gamma^e_{\rm dm} = \gamma_{\rm dm} - \frac{1}{\rho_{\rm dm}}
\left[ \zeta_0 \left(\frac{\rho_{\rm r0}}{a^4} + \frac{\rho_{\rm b0}}{a^3}
+ \rho_{\rm dm} \right) + \right. \\
\left. + \frac{\rho_{\rm de0}}{a^{3(\gamma_{\rm de} + \alpha)}}
(\alpha + \zeta_0 ) \right] \, ,
\end{multline}
and then the equation of motion~(\ref{ConsEqDE}) for the DM energy
density becomes
\begin{eqnarray}\label{ConsEqDM-2}
  \dot{\rho}_{\rm dm} &=& - 3H \left[ \gamma_{\rm dm} \rho_{\rm dm}
    \vphantom{\frac{\rho_{r0}}{a^4}} - \zeta_0 \left(\frac{\rho_{\rm
          r0}}{a^4} + \frac{\rho_{\rm b0}}{a^3} + \rho_{\rm dm}
    \right) \right. \nonumber \\
  && \left. - \frac{\rho_{\rm de0}}{a^{3(\gamma_{\rm de} + \alpha)}}
    (\alpha + \zeta_0 ) \right] \, .
\end{eqnarray}

\begin{widetext}
Next, we take the dimensionless density parameters for all matter
components, $\Omega_{i0} \equiv \rho_{i0}/\rho^0_{\rm crit}$ and
$\hat{\Omega}_{\rm dm} \equiv \rho_{\rm dm}/\rho^0_{\rm crit}$, where
$\rho_{\rm crit}^0 \equiv 3H^2_0/ (8 \pi G)$ is the present critical
density; thus, Eq.~ (\ref{ConsEqDM-2}) becomes
\begin{equation}
  \label{ConsEqDM-OmegaDMz}
  \frac{(1+z)}{3}\frac{d \hat{\Omega}_{\rm dm}}{dz} -
  \hat{\Omega}_{\rm dm} (\gamma_{\rm dm}-\zeta_0) +  \Omega_{\rm de0}
  (\alpha + \zeta_0) (1+z)^{3 ( \gamma_{\rm de} + \alpha)} + \zeta_0
  (1+z)^3 \left[ \Omega_{\rm r0} (1+z) + \Omega_{\rm b0} \right] =0
  \, . 
\end{equation}
where $z$ is the redshift, which is related to the scale factor
through $a=1/(1+z)$. The analytical solution of
Eq.~(\ref{ConsEqDM-OmegaDMz}) is:

\begin{multline}\label{SolutionConsEqDM-OmegaDM-AlphaNeqZeta0-Gdm1}
\hat{\Omega}_{\rm dm} (z) = \frac{1}{(1+z)^{3 \zeta_0} (1+3
\zeta_0)(\alpha + \gamma_{\rm de} + \zeta_0 -1)} \left\{ 3 \zeta_0^2 \left[
(\Omega_{\rm b0} + \Omega_{\rm dm0} + \Omega_{\rm r0} -1 ) (1+z)^{3(\alpha +
\gamma_{\rm de} + \zeta_0)} - \right. \right. \\
\left.  - (1+z)^{3(1+\zeta_0)} \left( (1+z) \Omega_{\rm
r0} + \Omega_{\rm b0} \right) + (1+z)^3 \right] + \alpha(3 \zeta_0 +
1)(\Omega_{\rm b0} + \Omega_{\rm dm0} + \Omega_{\rm r0} -1 ) (1+z)^{3(\alpha +
\gamma_{\rm de} + \zeta_0)} + \\
+ \zeta_0 \left[ (\Omega_{\rm b0} +
\Omega_{\rm dm0} + \Omega_{\rm r0} -1 ) (1+z)^{3(\alpha +
\gamma_{\rm de} + \zeta_0)} + (1+z)^{3(1+\zeta_0)} \left( (2- 3\gamma_{\rm
de}) \Omega_{\rm b0} - 3 (1+z) (\gamma_{\rm de}-1) \Omega_{\rm r0} \right)
+ \right. \\
\left. + (1+z)^3 \left( 3(\gamma_{\rm de} - 1) (\Omega_{\rm
b0} + \Omega_{\rm dm0}) + (3 \gamma_{\rm de} -4)
\Omega_{\rm r0} +1 \right) \right] + \alpha (1+z)^3 \left[ (1+z)^{3 \zeta_0}
\left( -3(1+z)\zeta_0 \Omega_{\rm r0} - \right. \right.  \\
\left. \left. \left.  - (1+3\zeta_0) \Omega_{\rm b0} \right) + 3
\zeta_0 - \Omega_{\rm r0} +1 \right] - (1+z)^3 (\gamma_{\rm de} - 1) \left[
\Omega_{\rm b0} \left( (1+z)^{3\zeta_0} -1 \right) - \Omega_{\rm dm0}
\right] \right\},
\end{multline}

\end{widetext}
where we have set $\gamma_{\rm dm} = 1$, and made use of the present
Friedmann constraint $\Omega_{\rm de0} = 1- \left( \Omega_{\rm r0}
  -\Omega_{\rm b0} - \Omega_{\rm dm0}  \right)$.

\begin{acknowledgments}
A.A. and Y.L. thanks PROMEP and CONACyT for support for a postdoctoral stay at
the Departamento de F\'isica of the Universidad de Guanajuato. This
work was partially supported by PIFI, PROMEP, DAIP-UG, CAIP-UG,
CONACyT M\'exico under grant 167335, and the Instituto Avanzado de
Cosmologia (IAC) collaboration.
\end{acknowledgments}

\bibliography{InteractingFluids}

\end{document}